\documentclass{aa}

\usepackage{mathtools}
\usepackage{BibDef}
\usepackage{longtable}
\usepackage{booktabs}
\usepackage[dvipsnames]{xcolor}
\usepackage{natbib}
\usepackage{booktabs,caption}
\usepackage[flushleft]{threeparttable}

\usepackage[breaklinks=true]{hyperref}

\usepackage[letterspace=150]{microtype}
\usepackage{kantlipsum}
\def \teff {$T_{\mathrm{eff}}$}
\newcommand{\juliet}{{\sc \tt juliet}\xspace}
\newcommand{\celerite}{{\sc \tt celerite}\xspace}
\newcommand{\batman}{{\sc \tt batman}\xspace}

\newcommand{\radvel}{{\sc \tt radvel}\xspace}
\newcommand{\dynesty}{{\sc \tt dynesty}\xspace}
\newcommand{\multinest}{{\sc \tt MultiNest}\xspace}
\newcommand{\spleaf}{{\sc \tt SPLEAF}\xspace}
\newcommand{\pymultinest}{{\sc \tt PyMultiNest}\xspace}
\newcommand{\cd}[1]{{\bf \textcolor{black}{#1}}}

\providecommand{\teff}{\ensuremath{T_{\rm eff}}}

\usepackage{graphicx}
\usepackage{caption}
\usepackage{subcaption}
\usepackage{txfonts}

\usepackage{caption}
\usepackage{subcaption}

\begin{document} 

\authorrunning{J. F. Otegi et al. }

   \title{{TESS and HARPS reveal two sub-Neptunes around TOI\,1062}  \vspace{8mm}}

  \author{J. F.~Otegi$^{1,2}$
  \and  F.~Bouchy$^{1}$
  \and  R.~Helled  $^{2}$
  \and  D.J.~Armstrong$^{19,20}$
  \and  M.~Stalport$^{1}$
  \and  K.G.~Stassun$^{13}$
  \and  E.~Delgado-Mena$^{17}$
  \and  N.C.~Santos$^{17,18}$
  \and  J.B.~Delisle$^{1}$
  \and  N.C.~Hara$^{1}$
  \and  K.~Collins$^{7}$
  \and  S. Gandhi $^{19,20}$
  \and  C.~Dorn$^{2}$
  \and  M.~Brogi$^{19,20}$
  \and  M.~Fridlund$^{24,25}$
  \and  H.P.~Osborn$^{23}$
  \and  S.~Hoyer$^{26}$
  \and  S.~Udry$^{1}$
  \and  S.~Hojjatpanah$^{17,18}$
  \and  L.D.~Nielsen $^{1}$
  \and  X.~Dumusque$^{1}$
  \and  V.~Adibekyan$^{17,18}$
  \and  D.~Conti$^{8}$
  \and  R.~Schwarz$^{9}$ 
  \and  G.~Wang$^{10}$
  \and  P.~Figueira $^{17,27}$
  \and  J.~Lillo-Box$^{3}$
  \and  R.F.~Díaz$^{28}$
  \and  A.~Hadjigeorghiou$^{19,20}$
  \and  D.~Bayliss$^{19,20}$
  \and  P.A.~Strøm$^{19,20}$
  \and  S.G.~Sousa$^{17}$
  \and  D.~Barrado$^{3}$
  \and  A.~Osborn$^{19,20}$
  \and  S.S.C.~Barros$^{17,18}$
  \and  D.J.A.~Brown$^{19,20}$
  \and  J.D.~Eastman $^{7}$
  \and  D.R.~Ciardi$^{16}$
  \and  A.~Vanderburg$^{15}$
  \and  R.F.~Goeke$^{4}$
  \and  N.M.~Guerrero $^{4}$
  \and  P.T.~Boyd$^{14}$
  \and  D.A.~Caldwell $^{12,21}$
  \and  C.E.~Henze$^{12}$
  \and  B.~McLean $^{22}$
    \and G.~Ricker$^{4}$
    \and R.~Vanderspek$^{4}$
    \and D.W.~Latham$^{7}$
    \and S.~Seager$^{4,5,6}$
    \and J.~Winn$^{11}$
    \and J.M.~Jenkins $^{12}$
}

  \institute{Observatoire Astronomique de l’Universit\'e de Gen\`eve, 51 Ch. des Maillettes, – Sauverny – 1290 Versoix, Switzerland\label{Geneva}
         \and Institute for Computational Science, University of Zurich,
              Winterthurerstr. 190, CH-8057 Zurich, Switzerland \label{Zurich}
          \and Center for Astrobiology (CAB, INTA-CSIC), Dept. de Astrof\'isica, ESAC campus 28692 Villanueva de la Ca\~nada (Madrid), Spain \label{Villanueva}
          \and Department of Physics and Kavli Institute for Astrophysics and Space Research, Massachusetts Institute of Technology, Cambridge, MA 02139, USA\label{MIT}
          \and Department of Earth, Atmospheric and Planetary Sciences, Massachusetts Institute of Technology, Cambridge, MA 02139, USA\label{MIT2}
        \and Department of Aeronautics and Astronautics, MIT, 77 Massachusetts Avenue, Cambridge, MA 02139, USA\label{MIT3}
        \and Center for Astrophysics \textbar \ Harvard \& Smithsonian, 60 Garden Street, Cambridge, MA 02138, USA\label{Harvard}
        \and American Association of Variable Star Observers, 49 Bay State Road, Cambridge, MA 02138, USA\label{AAVSO}
        \and Patashnick Voorheesville Observatory, Voorheesville, NY 12186, USA\label{Voorheesville}
        \and Tsinghua International School, Beijing 100084, China\label{Tsinghua}
        \and Department of Astrophysical Sciences, Princeton University, NJ 08544, USA\label{Princeton}
        \and NASA Ames Research Center, Moffett Field, CA 94035, USA\label{Ames}
        \and Vanderbilt University, Department of Physics \& Astronomy, Nashville, TN 37235, USA\label{Vanderbilt}
        \and Astrophysics Science Division, NASA Goddard Space Flight Center, Greenbelt, MD 20771, USA \label{Greenbelt}
        \and Department of Astronomy, The University of Texas at Austin, Austin, TX 78712, USA \label{Texas}
        \and Caltech/IPAC-NExScI, M/S 100-22, 1200 E California Blvd, Pasadena, CA 91125, USA \label{Caltech}
        \and Instituto de Astrof\'isica e Ci\^encias do Espa\c{c}o, Universidade do Porto, CAUP, Rua das Estrelas, 4150-762 Porto, Portugal\label{Oporto1}
        \and Departamento de F\'isica e Astronomia, Faculdade de Ci\^encias, Universidade do Porto, Rua do Campo Alegre, Porto, Portugal \label{Oporto2}
        \and Department of Physics, University of Warwick, Coventry CV4 7AL, UK\label{Warwick1}
        \and Centre for Exoplanets and Habitability, University of Warwick, Gibbet Hill Road, Coventry CV4 7AL, UK\label{Warwick2}
        \and SETI Institute, Mountain View, CA 94043, USA\label{seti}
        \and Space Telescope Science Institute, 3700 San Martin Drive, Baltimore, MD 21218, USA\label{Baltimore}
        \and NCCR/PlanetS, Centre for Space \& Habitability, University of Bern, Bern, Switzerland AND Department of Physics and Kavli Institute for Astrophysics and Space Research, Massachusetts Institute of Technology, Cambridge, MA 02139, USA\label{Bern}
        \and Leiden Observatory, University of Leiden, PO Box 9513, 2300 RA, Leiden, The Netherlands\label{Leiden}
        \and Department of Earth and Space Sciences, Chalmers University of
        Technology, Onsala Space Observatory, 439 92, Onsala, Sweden\label{Onsala}
        \and Aix Marseille Univ, CNRS, CNES, LAM, Marseille, France\label{Marseille}
        \and European Southern Observatory, Alonso de C\'{o}rdova 3107, Vitacura, Santiago, Chile\label{ESO}
        \and International Center for Advanced Studies (ICAS) and ICIFI (CONICET), ECyT-UNSAM, Campus Miguelete, 25 de Mayo y Francia, (1650) Buenos Aires, Argentina.\label{Buenos_Aires}}


  \abstract
  { The Transiting Exoplanet Survey Satellite (\textit{TESS}) mission was designed to perform an all-sky search of planets around bright and nearby stars. Here we report the discovery of two sub-Neptunes orbiting around the TOI\,1062 (TIC 299799658), a V=10.25 G9V star observed in the \textit{TESS} Sectors 1, 13, 27 \& 28. We use precise radial velocity observations from HARPS to confirm and characterize these two planets. TOI\,1062b has a radius of 2.265$^{+0.095}_{-0.091}$ R$_{\oplus}$, a mass of  11.8 $\pm 1.4$ M$_{\oplus}$, and an orbital period of 4.115050$\pm 0.000007$ days. The second planet is not transiting, has a minimum mass of 7.4$^{+1.6}_{-1.6}$ M$_{\oplus}$ and is near the 2:1 mean motion resonance with the innermost planet with an orbital period of 8.13$^{+0.02}_{-0.01}$ days. We performed a dynamical analysis to explore the proximity of the system to this resonance, and to attempt at further constraining the orbital parameters. The transiting planet has a mean density of 5.58$^{+1.00}_{-0.89}$ g cm$^{-3}$ and an analysis of its internal structure reveals that it is expected to have a small volatile envelope accounting for 0.35\% of the mass at maximum. The star's brightness and the proximity of the inner planet to the "radius gap" make it an interesting candidate for transmission spectroscopy, which could further constrain the composition and internal structure of TOI\,1062b. }
   \maketitle

\section{Introduction}

The Kepler mission was the first exoplanet mission to perform a large statistical survey of transiting exoplanets \cite[][]{Borucki-10,Howell-14}, and it impacted the field significantly with the detection of over 2300 exoplanets \cite[see NASA Exoplanet Archive][]{Akeson-13}. Nevertheless, due to the faintness of the stars targeted by Kepler, only a small fraction are suitable for radial velocity (RV) follow-up. The Transiting Exoplanet Survey Satellite (TESS) has been designed to survey 85\% of the sky for transiting exoplanets around bright, nearby stars \cite[][]{Ricker-15}. It spent the first year of its mission searching for planets in the Ecliptic Southern Hemisphere. More than 2000 candidates have been detected and over 80 have been confirmed, expanding the number of small planets around cool stars \cite[][]{Guerrero-20}. RV follow-up programs have allowed for a rapid  validation of transiting planet candidates and to constrain their masses. Accurate measurements of the planetary mass and radius provide information on the planet's bulk density. However, a full characterization on the planetary composition and internal structure is extremely challenging due to the degenerate nature of this problem, where various compositions and interiors can lead to the same average density. Nevertheless, by studying the population of planets in this mass/size range, it is possible to better understand the dominating processes of planet formation and evolution in a statistical manner \cite[][]{Alibert-10,Alibert-15}. \\

A key signature in the planet population is the ``hot Neptunian desert", which corresponds to a lack of Neptune-mass planets close to their host stars \cite[][]{Szabo-11}. The desert is likely to arise by a combination of photoevaporation and tidal disruption \cite[][]{Beauge-13,Mazeh2016,Owen-18}. At small orbital distances, planets are subject to intense radiation from the their host stars, and Neptune-like planets are probably not  massive enough to retain their gaseous envelopes. An important advantage of the all-sky nature of TESS resides in its ability to find the brightest examples of stars hosting rare sub-populations of planets, such as the exoplanets lying inside the desert. Indeed, TESS has been able to find the first planets deep inside the desert, for example TOI-849b \cite[][]{Armstrong2020}, a recently discovered remnant core of a giant planet, and LTT9779b \cite[][]{Jenkins-20}, an ultra-hot Neptune. The NCORES HARPS large program is designed to further study planets which may have undergone substantial envelope loss. Some of the recently discovered planets under this program include the mentioned TOI-849b and the three mini Neptunes TOI-125b, c, d  \cite[][]{Nielsen2020}.

Another relevant feature in the exoplanet population is a lack of planets with radii between $1.5 R_{\oplus}$ and $2 R_{\oplus}$ \cite[][]{Fulton2017} known as the "radius valley". This suggests that there is a transition between the super-Earth and sub-Neptune populations \cite[][]{Owen-12}. This gap could be a result of stellar irradiation \citep[][]{Lopez-13} or core-powered mass-loss \citep[e.g.,][]{Ginzburg-18} that leads to atmospheric evaporation that strips the planet down to its core or, alternatively, consequence of a gas-poor formation \citep[e.g.,][]{Gupta2019}. The observed features of the exoplanet populations are not well  understood and more exoplanet discoveries are crucial to reveal the overall picture of planets in the super-Earth and Neptune-size/mass  regimes.\\
\begin{figure*}[h]
\centering
  \begin{tabular}{@{}cc@{}}
    \includegraphics[width=1.0\textwidth]{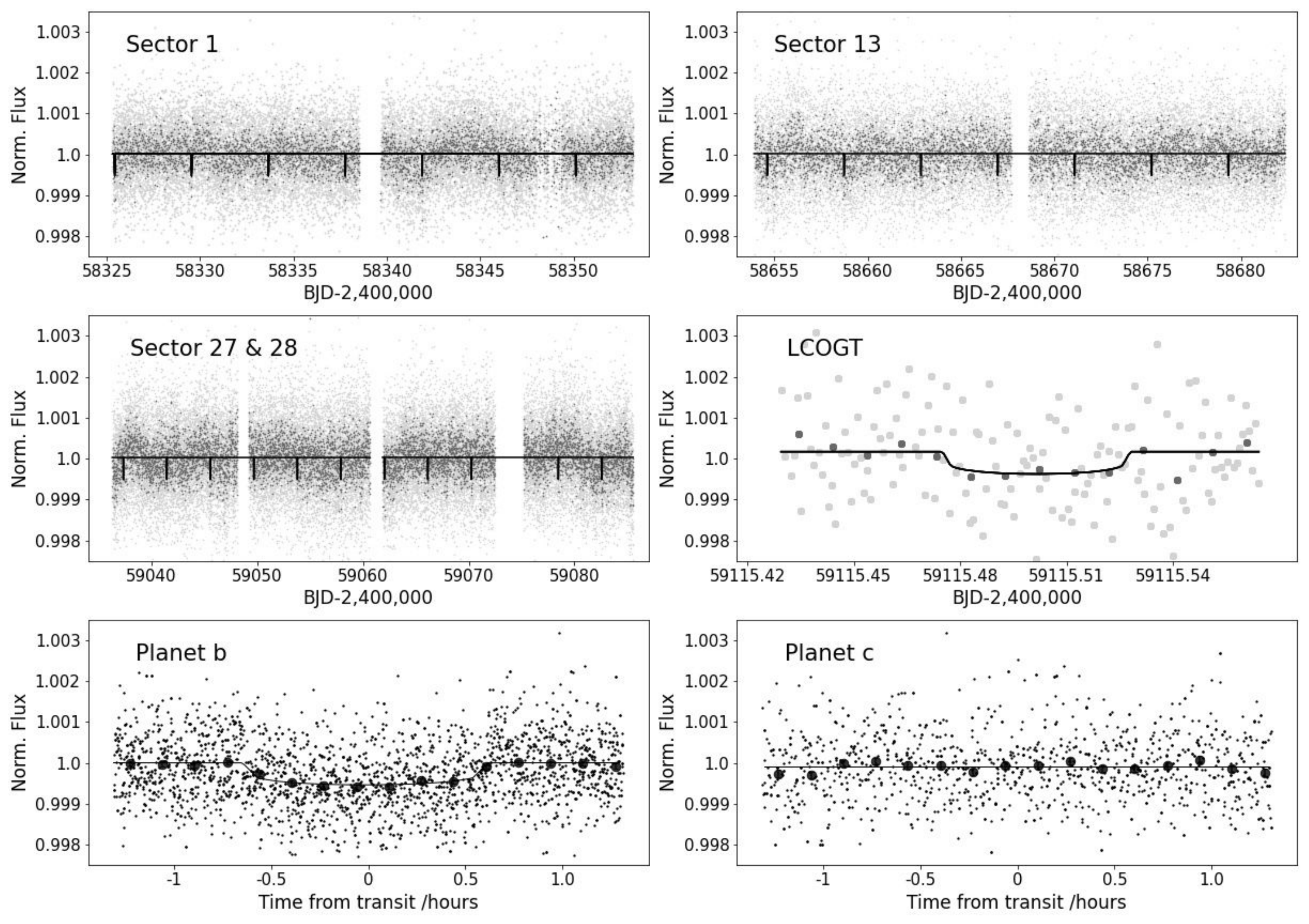}
  \end{tabular}
  \caption{TESS photometry from the of TOI\,1062 from Sectors 1 (upper left), 13 (upper right), and Sectors 27 \& 28 (middle left), and LCOGT photometry (middle right). The TESS light-curves correspond to the PDC-SAP flux time series provided by SPOC}. The TESS full light curve with the 2-minute cadence data is shown in light-grey, and the same data binned to 10 min in dark-grey. The gaps in the data coverage are due to observation interruptions of the TESS spacecraft to send the data to Earth, that occurs after each TESS orbit of 13.7 days. The dots of the LCOGT are bigger for better visualization. The lower panels shows the phase folded TESS light curves for TOI\,1062b (left) and TOI\,1062c (right) with the 2 min cadence data in light-grey  and binned to 10 minutes in black. We find that TOI\,1062c does not transit. \label{fig:light_curve} 
\end{figure*} 



\begin{table}
\caption{\label{tab:stellar} Stellar parameters of TOI-1062.}
\resizebox{\columnwidth}{!}{%
	\begin{tabular}{lcc}
	\hline\hline
	\noalign{\smallskip}
	Parameter	&	Value	&	Source\\

	\hline
    \noalign{\smallskip}
    \noalign{\smallskip}
    \multicolumn{3}{l}{\underline{Identifying Information}}\\
    \noalign{\smallskip}
    \noalign{\smallskip}
    TOI & TOI-1062 &  TESS \\
    TIC  ID & 299799658 &  TESS \\
    2MASS ID	&  J02322848-7801256	& 2MASS \\
    \multicolumn{2}{l}{Gaia ID ~~~~~~~~~~~~~~~~~~~~~~~~~~~4632865331094140928}	& Gaia DR3 \\

    \\
    \multicolumn{3}{l}{\underline{Astrometric Parameters}}\\
    \noalign{\smallskip}
    \noalign{\smallskip}
    R.A. (J2015.5, h:m:s)		&	 02:32:29 	&  Gaia DR3	\\
	Dec	 (J2015.5, h:m:s)		&	 -78:01:25	&  Gaia DR3	\\
    Parallax  (mas) & 12.14 $\pm$0.01 & Gaia DR3\\
    Distance  (pc) & 82.2$\pm$0.2 & Gaia DR3\\
    \\
    \multicolumn{3}{l}{\underline{Photometric Parameters}}\\
    \noalign{\smallskip}
    \noalign{\smallskip}
	B 		&11.02 $\pm$ 0.07  	&Tycho \\
	V 		&10.25 $\pm$ 0.01	&Tycho\\
	T 	    & 9.48 $\pm$ 0.01	& TESS\\
    G 		& 10.00 $\pm$ 0.01	&{Gaia}\\
    J 		& 8.78  $\pm$ 0.02	&2MASS\\
   	H 		&  8.41 $\pm$ 0.03	&2MASS\\
	K      & 8.30$\pm$ 0.03	&2MASS\\
    W1  & 8.257 $\pm$ 0.02	&WISE\\
    W2  & 8.322 $\pm$ 0.02	&WISE\\
    W3  & 8.256  $\pm$ 0.02 & WISE\\
    W4  & 8.35 $\pm$ 0.19 & WISE \\
    A$_{V}$	& 0.13 $\pm$ 0.04 & Sec. \ref{sec:star} \\
    
    \\
    
    \multicolumn{3}{l}{\underline{Abundances}}\\
    \noalign{\smallskip}
    \noalign{\smallskip}
	$\mathrm{\left[Fe/H\right](dex)}$  & 0.14 $\pm$ 0.04 &Sec. \ref{sec:star} \\
	$\mathrm{\left[O/H\right](dex)}$  & -0.14 $\pm$ 0.16 &Sec. \ref{sec:star} \\
	$\mathrm{\left[C/H\right](dex)}$  & 0.13 $\pm$ 0.02 &Sec. \ref{sec:star} \\
	$\mathrm{\left[Cu/H\right](dex)}$  & 0.25 $\pm$ 0.04 &Sec. \ref{sec:star} \\
	$\mathrm{\left[Zn/H\right](dex)}$  & 0.08 $\pm$ 0.04 &Sec. \ref{sec:star} \\
	$\mathrm{\left[Sr/H\right](dex)}$  & 0.17 $\pm$ 0.08 &Sec. \ref{sec:star} \\
	$\mathrm{\left[Y/H\right](dex)}$  & 0.04 $\pm$ 0.08 &Sec. \ref{sec:star} \\
	$\mathrm{\left[Zr/H\right](dex)}$  & 0.20 $\pm$ 0.06 &Sec. \ref{sec:star} \\
	$\mathrm{\left[Ba/H\right](dex)}$  & -0.02 $\pm$ 0.04 &Sec. \ref{sec:star} \\
	$\mathrm{\left[Ce/H\right](dex)}$  & 0.16 $\pm$ 0.10 &Sec. \ref{sec:star} \\
	$\mathrm{\left[Nd/H\right](dex)}$  & 0.10 $\pm$ 0.06 &Sec. \ref{sec:star} \\
	$\mathrm{\left[MgI/H\right](dex)}$  & 0.16 $\pm$ 0.07 &Sec. \ref{sec:star} \\
	$\mathrm{\left[AlI/H\right](dex)}$  & 0.25 $\pm$ 0.04 &Sec. \ref{sec:star} \\
	$\mathrm{\left[SiI/H\right](dex)}$  & 0.20 $\pm$ 0.07 &Sec. \ref{sec:star} \\
	$\mathrm{\left[CaI/H\right](dex)}$  & 0.10 $\pm$ 0.08 &Sec. \ref{sec:star} \\
	$\mathrm{\left[TiI/H\right](dex)}$  & 0.23 $\pm$ 0.07 &Sec. \ref{sec:star} \\
	$\mathrm{\left[CrI/H\right](dex)}$  & 0.17 $\pm$ 0.07 &Sec. \ref{sec:star} \\
	$\mathrm{\left[NiI/H\right](dex)}$  & 0.17 $\pm$ 0.05 &Sec. \ref{sec:star} \\

\\

    \\

  \multicolumn{3}{l}{\underline{Bulk Parameters}}\\
    \noalign{\smallskip}
    \noalign{\smallskip}
    Mass ($M_{\odot}$)                   & 0.94 $\pm$ 0.02  & Sec. \ref{sec:star}\\
    Radius ($R_{\odot}$)                 & 0.84 $\pm$ 0.09  & Sec. \ref{sec:star}\\
    \teff\,(K)                           & 5328 $\pm$ 56    & Sec. \ref{sec:star}\\
    log g (cm\,s$^{-2}$)                 & 4.55 $\pm$ 0.09  & Sec. \ref{sec:star}\\
    Spectral type                        & G9V              & Sec. \ref{sec:star}\\
    $\rho$ (g\,cm$^{-3}$)                & 2.22 $\pm$ 0.10  & Sec. \ref{sec:star}\\
	{\it v}\,sin\,{\it i} (km\,s$^{-1}$) & 2.13 $\pm$ 0.5	& Sec. \ref{sec:star}\\
	P$_{rot}$ (d)                        & 21.8 $\pm$ 2.4	&Sec. \ref{sec:star} \\
	Age	(Gyrs)                           & 2.5 $\pm$ 0.3	&Sec. \ref{sec:star} \\

 	\noalign{\smallskip}
	\hline
 	\noalign{\smallskip}
    \end{tabular}}
    
 2MASS \citep{2MASS};Tycho \citep{Tycho}; WISE \citep{WISE}; Gaia \citep{Gaia2018} 
\end{table}       

 In this paper we present the discovery and confirmation of two highly irradiated sub-Neptunes hosted by the TESS Object of Interest (TOI) 1062, a nearby (d= 82 pc) and bright ($V_{mag}$ \~ 10.2) G9V star (see Table 1 for a full summary of the stellar properties).  We also used intensive radial velocity follow-up observations with HARPS to confirm the planetary nature of the transit
detected in the TESS data and precisely determine the properties of the planetary system. The paper is organized as follows: in Section 2 we present the data collected on the system to discover and
validate the planets, in Section 3 we analyze the host star fundamental parameters, and we characterize and discuss the planets in Section 4. Our conclusions are presented in Section 5.  \\

\vspace{5mm}

\subsection{TESS photometry}

TESS observed TOI\,1062 (TIC 299799658) in Sectors 1 and 13 during the first year, and during Sectors 27 and 28 of the extended mission, obtaining data from 25th July to 22nd August 2018, from 19th June to 17th July 2018, and from 4th July to 26 August 2020. The two-minute cadence data were reduced with the Science Processing Operations Center (SPOC) pipeline \cite[][]{Jenkins-16} adapted from the pipeline for the Kepler mission at the NASA Ames Research Center in order to produce calibrated pixels and light curves. On 2018, September 29 a transit candidate was announced around TOI\,1062. TOI\,1062.01 is a planet candidate with a period of 4.11506$\pm$0.00002 days, a transit depth of 487.90 $\pm$ 45 ppm, and an estimated planet radius of 2.32 $\pm$ 0.46 $R_{\oplus}$. The candidate passed all the tests from the Threshold Crossing Event (TCE) Data Validation Report \citep[DVR;][]{Twicken-18,Li-19}. \\


We used the publicly available Presearch Data Conditioning (PDC-SAP) flux time series \citep[e.g.][]{Twicken-10,Smith-12} provided by SPOC for the transit modelling, which is corrected for common trends and artifacts, for crowding and for the finite flux fraction in the photometric aperture.  Figure \ref{fig:light_curve} shows the 2 min cadence TESS light curve, along with the phase folded curve for TOI\,1062b.

\begin{figure}[h]
  \begin{center}
    \includegraphics[scale=0.6]{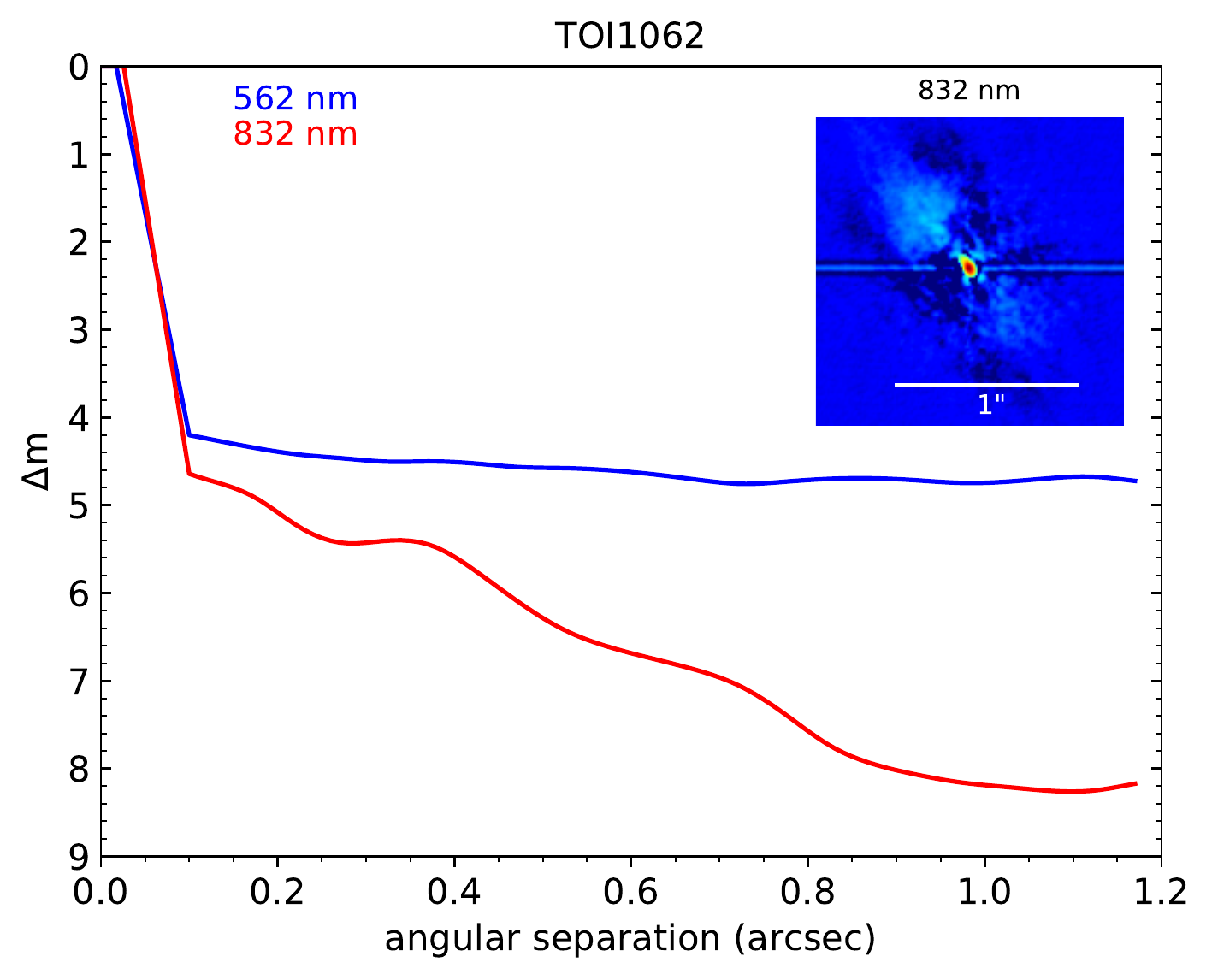}
    \caption{Gemini/Zorro contrast curves and 1.2" x 1.2" reconstructed image of the 832\,nm band.   \label{fig:zorro} }
 
  \end{center}
\end{figure}

\subsection{Ground based photometry with LCOGT}

We obtained ground-based time-series follow-up photometry of full transits of TOI\,1062.01 on 24th August 2019 in $i'$-band and on 22nd September 2020 in Pan-STARSS $z$-short band from the Las Cumbres Observatory Global Telescope \citep[LCOGT;][]{Brown-13} 1.0\,m network node at South Africa Astronomical Observatory. We used the TESS Transit Finder which is a customized version of the Tapir software package \cite[][]{jensen-13}, to schedule the observations. The 4096x4096 LCO SINISTRO cameras have an image scale of 0.389" per pixel, providing a field of view of 26'x26'. The standard LCOGT BANZAI pipeline was used to calibrate the images, and the photometric data were extracted with the AstroImageJ (AIJ) software package \cite[][]{Collins-17}.

The initial $i'$-band observation intentionally saturated the target star to check for a faint nearby eclipsing binary (NEB) that could be contaminating the \textit{TESS} photometric aperture. To account for possible contamination from the wings of neighboring star PSFs, we searched for NEBs out to 2.5' from the target star. If fully blended in the SPOC aperture, a neighboring star that is fainter than the target star by 8.8 magnitudes in TESS-band could produce the SPOC-reported flux deficit at mid-transit (assuming a 100\% eclipse). To account for possible delta-magnitude differences between TESS-band and $i'$-band, we included an extra 0.5 magnitudes fainter (down to \textit{TESS}-band magnitude 18.3). Our search ruled out NEBs in all 24 neighboring stars that meet our search criteria.  The $z$-short band observation was defocused and exposed appropriately to measure the TOI\,1062.01 light curve at high precision. For the data  reduction process we used a 10.5”-radius aperture and selected four comparison stars with brightness comparable to that of the target star. The inner and outer radii of the background annulus were then set to the corresponding AIJ-calculated values of 18.7” and 28”, respectively. This combination was chosen as it minimized the amount of noise in the light curve. A 0.5 ppt transit, with model residuals of 0.3 ppt in 10-minute bins, was detected in the uncontaminated target aperture.



 \begin{figure}[h]
  \begin{center}
    \includegraphics[width=0.75\linewidth,trim=70 80 80 80,clip,angle=90]{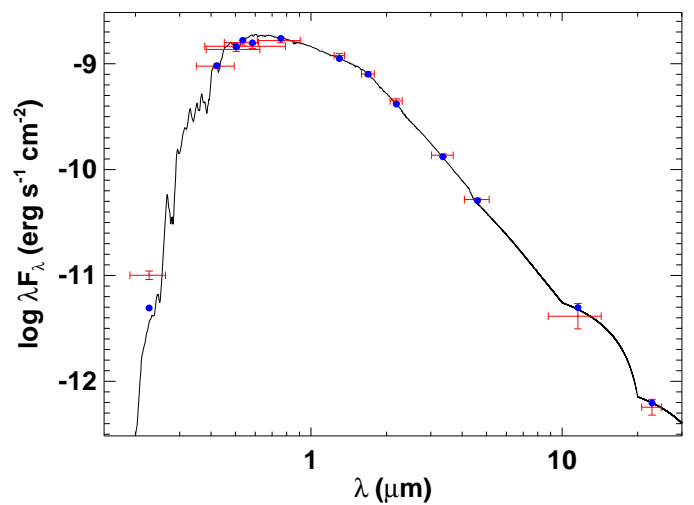}
    \caption{Spectral energy distribution of TOI\,1062. Red symbols represent the observed photometric measurements, where the horizontal bars represent the effective width of the passband. Blue symbols are the model fluxes from the best-fit Kurucz atmosphere model (black).  \label{fig:sed} }
 
  \end{center}
\end{figure}

\subsection{HARPS follow-up}

We collected 90 high-resolution spectra of TOI\,1062 using the High Accuracy Radial Velocity Searcher (HARPS) spectrograph mounted at the ESO 3.6m telescope of La Silla Observatory, Chile \citep[][]{Mayor-13}, with the goal of precisely determining the mass of the planet candidate and to search for additional planets in the system. The observations were carried out as part of the NCORES large programme (ID 1102.C-0249, PI: Armstrong) between 31st August 2019 and 17th January 2020. HARPS is a stabilised high-resolution (R~115000) echelle spectrograph which can reach sub-$m\,s^{-1}$ RV precision \cite[][]{Mayor-03}. The instrument was used in high-accuracy mode (HAM), with a 1" fiber on the star and another one to monitor the sky-background. We used exposure times of 30 minutes.

The standard HARPS Data Reduction Software (DRS) was used to reduce the data, using a K0  mask for both the cross-correlation function (CCF) \cite[][]{Pepe-02,Baranne-96} and the colour correction, and reaching a typical signal-to-noise ratio per pixel of 60 and a photon-noise uncertainty of 1.4 m s$^{-1}$ .  For each spectrum the usual activity indicators (S-index, H$\alpha$-index, Na-index, Ca-index, $\log R^{\prime}_\mathrm{HK}$), the full width half maximum (FWHM), the line bisector and the contrast of the CCF were measured. \\

\subsection{High resolution imaging}

 \begin{figure}[h]
  \begin{center}
    \includegraphics[scale=0.5]{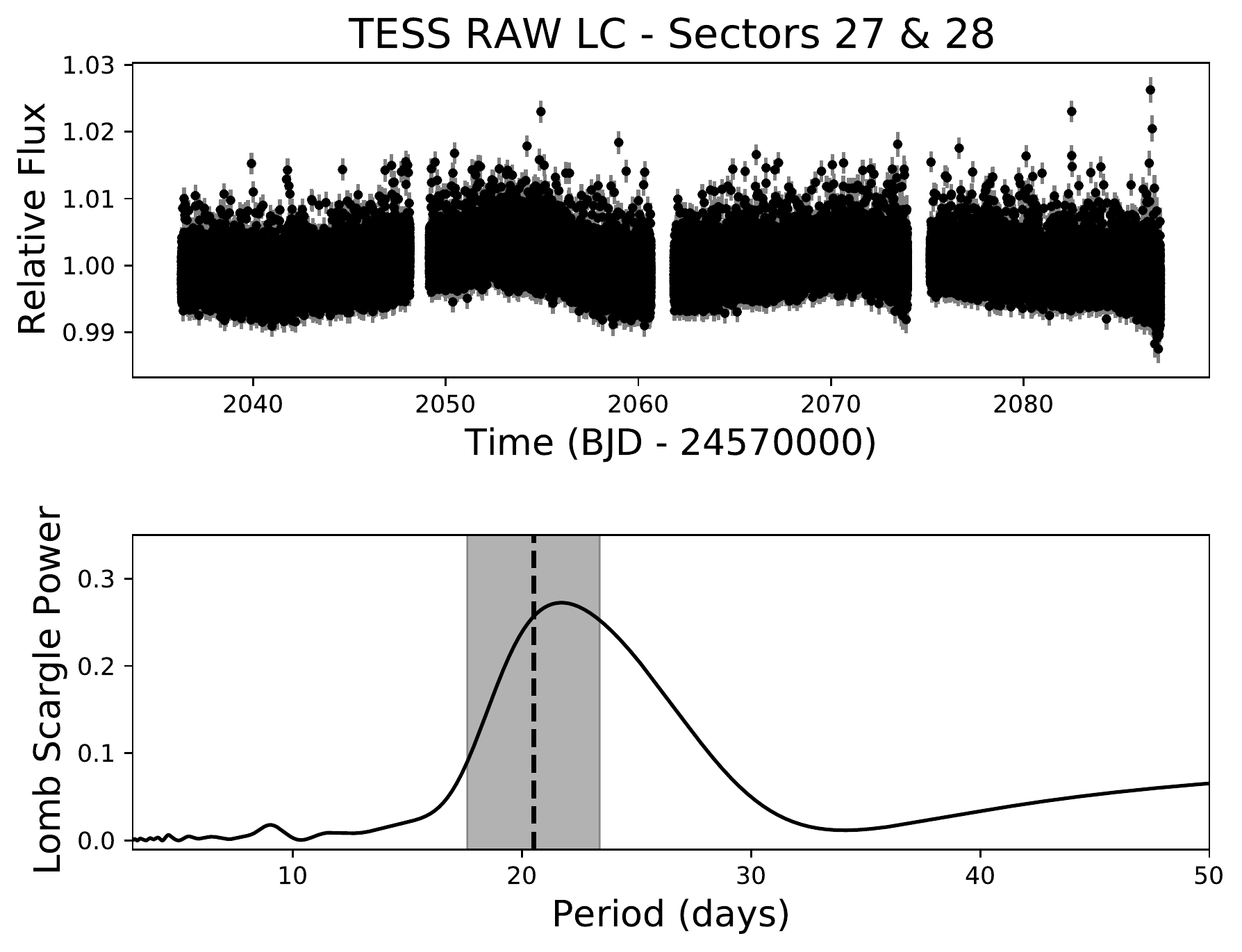}
    \caption{TESS RAW photometric data of sectors 27 and 28 (top) and the corresponding periodogram. The peak of the periodogram shows a possible rotational modulation with a period of 21.8 days, which is in agreement with the rotational period obtained from HARPS calibration. \label{fig:raw} }
 
  \end{center}
\end{figure}

To check for the presence of contaminating stars in the TESS photometric aperture, on 14th January 2020 TOI\,1062 was observed using the Zorro speckle imager \citep{Scott-19}, mounted on the 8.1m Gemini South telescope in Cerro Pachon, Chile. Zorro uses high speed electron-multiplying CCDs (EMCCDs) to simultaneously acquire data in two bands centered at 562\,nm and 832\,nm. The data is reduced following the procedures described in \cite{Howell-11}, and provides output data including a reconstructed image and robust limits on companion detections. The contrast achieved in the resulting reconstructed image is shown in Figure \ref{fig:zorro}. We see that the Zorro speckle images reach a contrast of $\Delta$mag=8.25 at a separation of 1.1" in the 832\,nm band, and cover a spatial range of 9.9 to 98.6 au around the star with contrasts between 5 and 8 mag in both bands, showing no close companions to TOI\,1062b.

\section{Host star fundamental parameters}
\label{sec:star}

\subsection{Analysis of the spectrum}

The stellar atmospheric parameters ($T_{\mathrm{eff}}$, $\log g$ and [Fe/H]) and respective error bars were derived using the methodology described in \citet[][]{Sousa-14, Santos-13}. Briefly, we make use of the equivalent widths (EW) of iron lines, as measured in the combined HARPS spectrum of TOI\,1062 using the ARES v2 code\footnote{The most recent version of ARES code (ARES v2) can be downloaded at http://www.astro.up.pt/$\sim$sousasag/ares} \citep{Sousa-15}, and we assume ionization and excitation equilibrium. This method makes use of a grid of Kurucz model atmospheres \citep{Kurucz-93} and the radiative transfer code MOOG \citep{Sneden-73}. This approach leads to the following stellar parameters: $T_{\mathrm{eff}}$= 5328 $\pm$ 56K,  [Fe/H]=0.14 $\pm$ 0.04 dex and $\log g$= 4.55 $\pm$ 0.09.  This is in agreement with the $\log g$ derived making use of the {\it Gaia} parallax and luminosity \citep[][]{Santos-04} which gives a value of 4.53. Using the \cite{Torres:2010} calibration leads to stellar mass of 0.94$\pm$0.02 $M_{\odot}$ and radius of 0.84$\pm$0.09 $R_{\odot}$. These values are used as priors in the global fit with \juliet in Section 4. \\

\subsection{Analysis of the spectral energy distribution}

As an independent determination of the stellar parameters, we also performed an analysis of the broadband spectral energy distribution (SED) of the star together with the {\it Gaia\/} DR2 parallax \citep[adjusted by $+0.08$~mas to account for the systematic offset reported by][]{StassunTorres:2018}, in order to determine an empirical measurement of the stellar radius, following the procedures described in \citet{Stassun:2016,Stassun:2017,Stassun:2018}. We pulled the $B_T V_T$ magnitudes from {\it Tycho-2}, the $BVi$ magnitudes from {\it APASS}, the $JHK_S$ magnitudes from {\it 2MASS}, the W1--W4 magnitudes from {\it WISE}, the $G, \,G_{\rm BP}, \,G_{\rm RP}$ magnitudes from {\it Gaia}, and the NUV magnitude from {\it GALEX}. Together, the available photometry spans the full stellar SED over the wavelength range 0.2--22~$\mu$m (see Figure~\ref{fig:sed}).  \\

We performed a fit using Kurucz stellar atmosphere models, with the effective temperature ($T_{\rm eff}$), metallicity ([Fe/H]), and surface gravity ($\log g$) adopted from the spectroscopic analysis. The only additional free parameter is the extinction ($A_V$), which we fixed to be zero due to the star's proximity. The resulting fit is very good (Figure \ref{fig:sed}) with a reduced $\chi^2$ of 1.8. The reduced $\chi^2$ is improved to 1.2 by excluding the GALEX NUV flux, which exhibits a modest excess; using the empirical relations of \citet{Findeisen:2011}, the GALEX NUV excess implies an activity level $\log R'_{\rm HK} = -4.9 \pm 0.1$, consistent with the spectroscopically determined value.

Integrating the (unreddened) model SED gives the bolometric flux at Earth, $F_{\rm bol} = 2.429 \pm 0.057 \times 10^{-9}$ erg~s$^{-1}$~cm$^{-2}$. Taking the $F_{\rm bol}$ and $T_{\rm eff}$ together with the {\it Gaia\/} DR2 parallax, gives the stellar radius, $R_\star = 0.838 \pm 0.020\,R_\odot$. In addition, we can use the $R_\star$ together with the spectroscopic $\log g$ to obtain an empirical mass estimate of $M_\star = 0.91 \pm 0.19 M_\odot$, which is consistent with that obtained via empirical relations of \citet{Torres:2010} and a 6\% error from the empirical relation itself, $M_\star = 0.97 \pm 0.06 M_\odot$. These values are in agreement with the ones derived from the spectral analysis in Section 3.1.   \\

\subsection{Rotational period and age}

\begin{figure}[h]
     \centering
     \begin{subfigure}[b]{0.5\textwidth}
         \centering
         \includegraphics[scale=0.47]{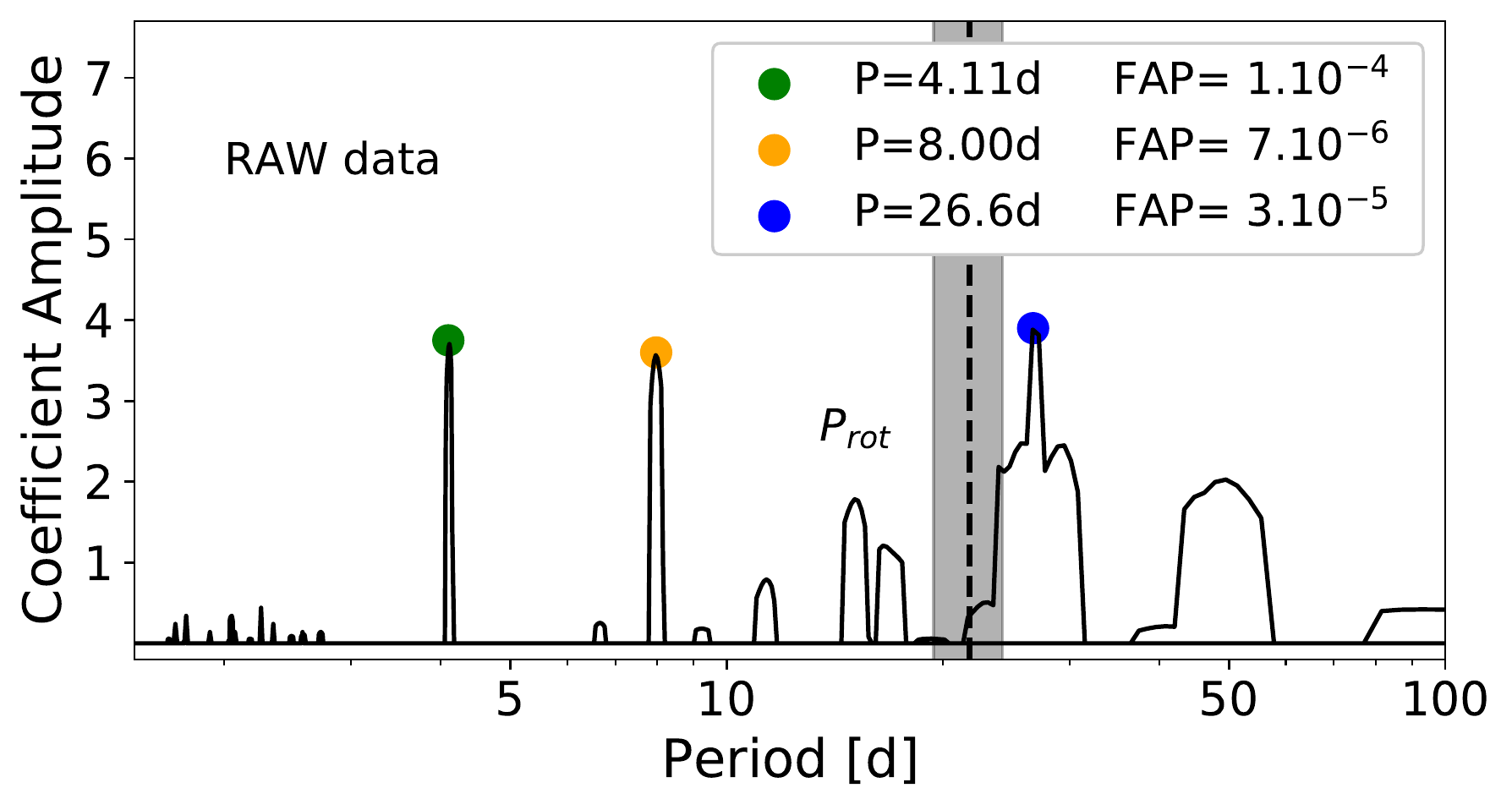}
     \end{subfigure}
     \hfill
     \begin{subfigure}[b]{0.5\textwidth}
         \centering
         \includegraphics[scale=0.47]{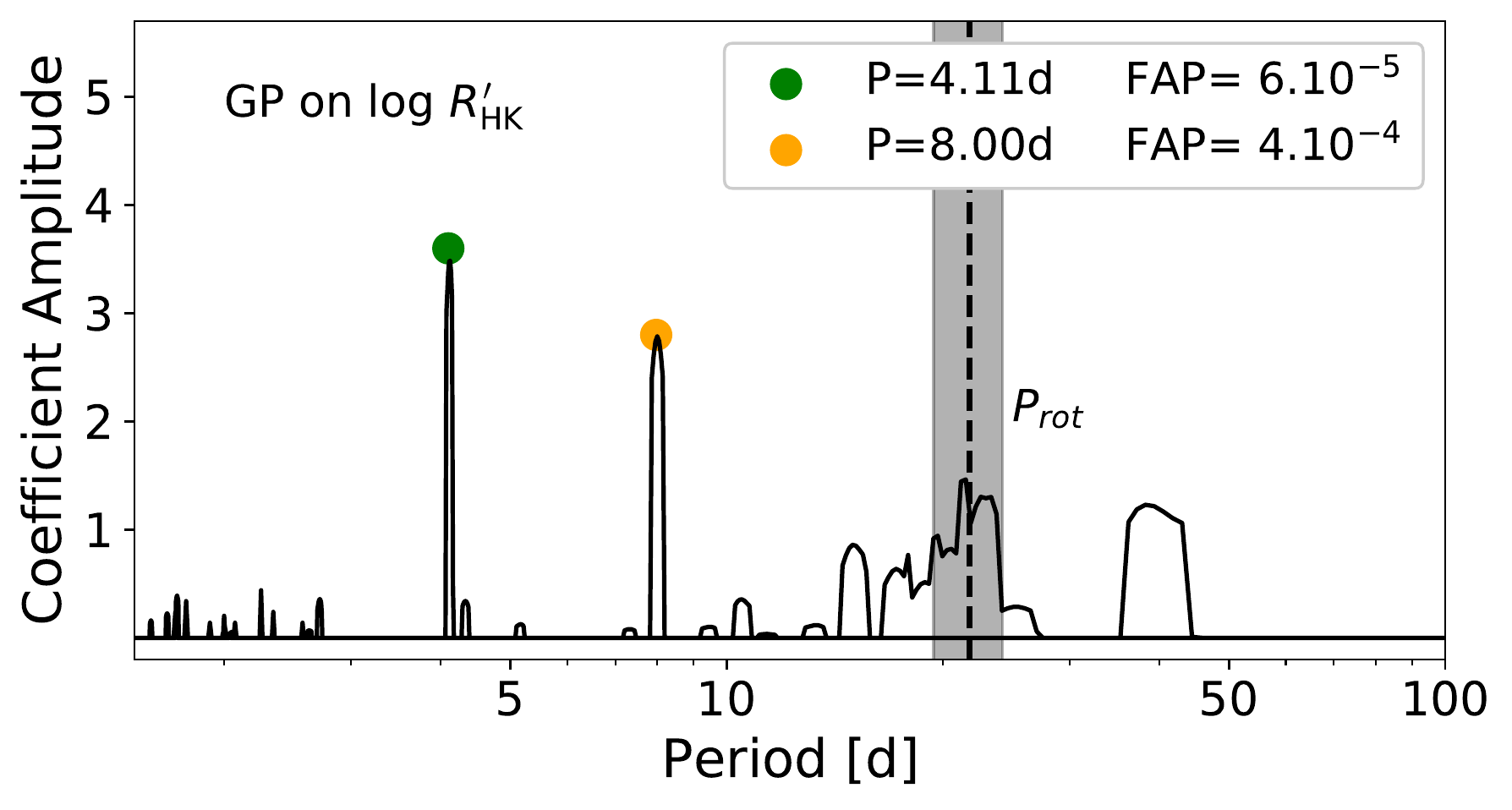}
     \end{subfigure}
        \caption{$l_1$ periodogram of the undetrended RVs (top) and the detrended RVs (bottom)  using gaussian processes on the $\log R^{\prime}_\mathrm{HK}$ indicator with \spleaf.The periods at which the significant peaks occur (FAP<$10^{-3}$) are represented in colors. The stellar rotational period determined through the TESS data is highlighted in grey.} \label{residuals} 
\end{figure}

\begin{figure}[h]
  \begin{center}
    \includegraphics[scale=0.48]{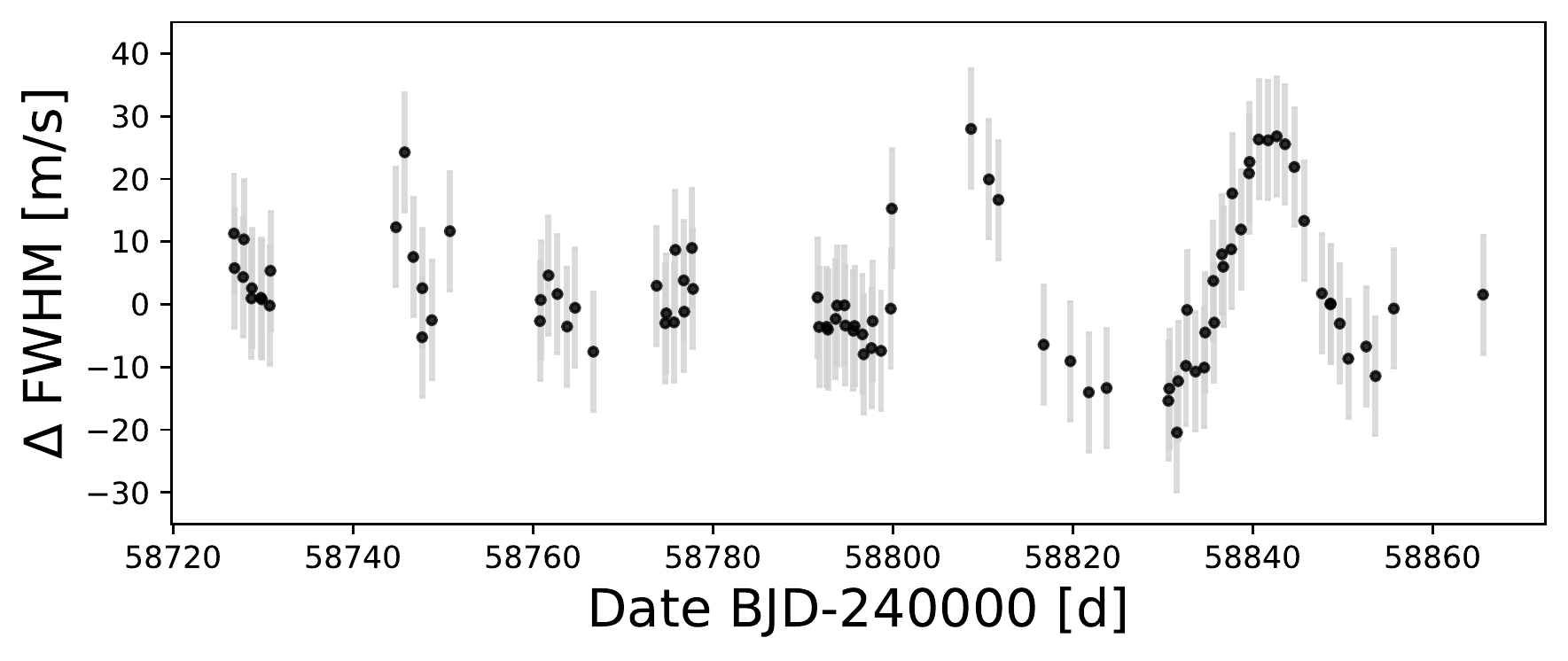}
    \caption{ Time-series of the CCF-FWHM.  \label{fig:fwhm} }
 
  \end{center}
\end{figure}

The stellar age can be  derived from empirical rotation-age relations of \citet{MamajekHillenbrand:2008}.  The HARPS calibration from the CCF-FWHM and the color index B-V gives $v\sin i$= 2.13 $\pm$ 0.5 $km/s$, then  together with $R_\star$, we estimate the stellar rotation period $P_{\rm rot}/\sin i = 20.5 \pm 2.9$~d. In addition, Fig \ref{fig:raw} shows that the peak of the periodogram of the TESS light-curves of Sectors 27 and 28 is at 21.8$2.4$ days. The measured rotational modulation is nearly identical in the other TESS sectors.  Another estimate of the expected rotation can be inferred from the average value of the Ca {\sc ii} H\,\&\,K chromospheric activity indicator for TOI\,1062, which is $\log R^{\prime}_\mathrm{HK}=-4.80\pm0.05$. Using the empirical relations in \citet{MamajekHillenbrand:2008} we get a rotation period of 21.9$^{+3}_{-2}$ days, in agreement with the result derived from the HARPS calibration and the modulation of the RAW lightcurves.  Another way of constraining the rotational period of the star consists of fitting the RVs with time-dependent gaussian processes (GP). We use the \celerite package \citep[e.g.,][]{foreman2017} with a quasi-periodic kernel, and obtain a rotational period of $22.4^{+4.6}_{-6.7}$ days, which is in agreement with the previous estimations.  \\

Using the rotational period obtained from the TESS data and the relations of \citet{MamajekHillenbrand:2008}, the age estimate is $\tau_\star$=2.5$\pm$0.3\,Gyr.  Note that, because the $v\sin i$ is a projected rotational velocity, it is a lower limit on the true rotational velocity, so $P_{\rm rot}/\sin i$ is an upper limit on the true $P_{\rm rot}$, which implies that the age estimate above provides an upper limit to the stellar age. \\

We also attempted to derive the stellar age by making use of the so-called chemical clocks \cite[e.g.][]{Delgado-19}, however we obtain a value much higher than with other methods described below. The reason is that the low $T_{\rm eff}$ of TOI\,1062 lies at the limit of applicability of such empirical relations and thus we refrained to use the age provided by this method.

\begin{table}
\caption{\label{tab:priors} Prior parameter distribution of the global fit with \juliet. $\mathcal{U}(a,b)$ indicates a uniform distribution between $a$ and $b$; $\mathcal{J}(a,b)$ a Jeffrey or log-uniform distribution between $a$ and $b$; and $\mathcal{N}(a,b)$ a normal distribution with mean $a$ and stardard deviation $b$.}
\resizebox{\columnwidth}{!}{%
	\begin{tabular}{lcccc}
	\hline\hline
	\noalign{\smallskip}
	Parameter	&&&&	Prior distribution	\\

	\hline
    \noalign{\smallskip}
    \noalign{\smallskip}
    \multicolumn{2}{l}{\underline{Instrumental Parameters:}}&&&\\
    \noalign{\smallskip}
    \noalign{\smallskip}
    ~~~~$q_{1,TESS}$  & &&& $\mathcal{U}(0,1)$ \\
    ~~~~$q_{2,TESS}$  & &&& $\mathcal{U}(0,1)$  \\
    ~~~~$m_{flux,TESS}$  & &&&  $\mathcal{N}(0,0.1)$	\\
    ~~~~$\sigma _{TESS}$  & [ppm] &&&  \cd{$\mathcal{J}(0.1,1000)$}	\\
    ~~~~$\sigma _{HARPS}$  & [m/s] &&&  $\mathcal{J}(0.1,1000)$	\\
    
    \noalign{\smallskip}
    \noalign{\smallskip}
    \multicolumn{2}{l}{\underline{GP parameters:}}&&&\\
    \noalign{\smallskip}
    \noalign{\smallskip}
    ~~~~$GP_{\sigma}$  & [m/s] &&& $\mathcal{J}(10^{-3},10^5)$ \\
    ~~~~$GP_{\rho}$  &&&& $\mathcal{J}(10^{-3},10^5)$ \\
    ~~~~$GP_T$  & [d] &&& $\mathcal{J}(10^{-3},10^5)$\\

        \noalign{\smallskip}
    \noalign{\smallskip}
    \multicolumn{2}{l}{\underline{Planetary parameters:}}&&&\\
    \noalign{\smallskip}
    \noalign{\smallskip}
    ~~~~$P_b$  & [d] &&& $\mathcal{N}(4.11,0.1)$ \\
    ~~~~$P_c$  & [d] &&& $\mathcal{N}(8.00,0.1)$ \\
    ~~~~$T_{0,b}$  & [$BJD_{TBD}$] &&& $\mathcal{N}(2459082.59485,0.1)$ \\
    ~~~~$T_{0,c} ^*$  & [$BJD_{TBD}$]  &&& $\mathcal{U}(2459082,2459091)$ \\
    ~~~~$K_b$  & [m/s] &&& $\mathcal{U}(1,100)$ \\
    ~~~~$K_c$  & [m/s] &&& $\mathcal{U}(1,100)$ \\
    ~~~~$\sqrt{e}$ sin($\omega$)  &  &&& $\mathcal{U}(-1,1)$ \\
    ~~~~$\sqrt{e}$ cos($\omega$)  &  &&& $\mathcal{U}(-1,1)$ \\
    ~~~~$r_1$&  &&& $\mathcal{U}(0,1)$ \\
    ~~~~$r_2$&  &&& $\mathcal{U}(0,1)$ \\

 	\noalign{\smallskip}
	\hline
 	\noalign{\smallskip}

    \end{tabular}
    
    }

\end{table}

\subsection{Stellar abundances}

Stellar abundances of refractory elements were derived using the classical curve-of-growth analysis method assuming local thermodynamic equilibrium \citep[e.g.][]{Adibekyan-12, Adibekyan-15, Delgado-17}.  Abundances of the volatile elements, C and O, were derived following the method of \cite{Delgado-10, Bertrandelis-15}. Since the two spectral lines of oxygen are usually weak and the 6300.3\AA{} line is blended with Ni and CN lines, the EWs of these lines were manually measured with the task \texttt{splot} in IRAF. All the [X/H] ratios are obtained by doing a differential analysis with respect to a high S/N solar (Vesta) spectrum from HARPS. The obtained values for this star are normal considering its metallicity except for the lower than expected value of oxygen. We find a high dispersion between both oxygen indicators that has to be taken with caution due to the difficulty of measuring reliable oxygen abundances for cool metal rich stars. The stellar abundances of the elements are presented in Table \ref{tab:stellar}. 

\subsection{Signal identification}
\label{sec:signals}

\begin{figure*}[h]
\centering
  \begin{tabular}{@{}cc@{}}
    \includegraphics[scale=0.57]{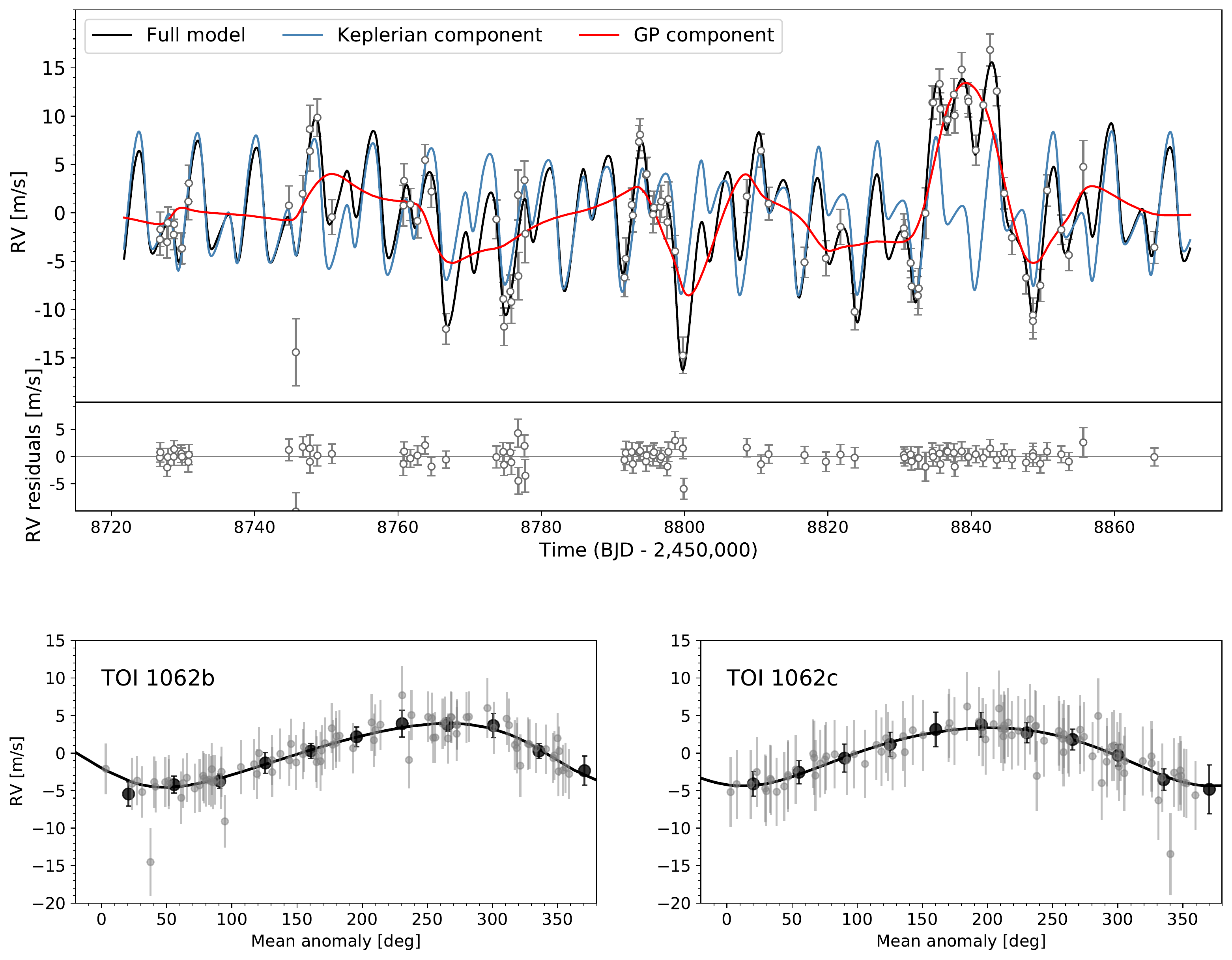}
  \end{tabular}
  \caption{The top panel shows the HARPS RVs for TOI\,1062 with the two planet model including gaussian processes and eccentric orbit, along with the residuals to the best fit right below the timeseries.} The bottom panel shows the phase folded data for each planet, with the gray dots representing the observed RVs and the black dots the binned data.  \label{rvs}
\end{figure*}

In order to search for periodicities in the RV data, we computed the $l_1$  periodogram \citep[e.g.][]{Hara-17,Hara-20}. This tool is  designed to find periodicities in time-series data corresponding to exoplanets detected using radial velocity data. It can be used similarly to a typical Lomb-Scargle periodogram or its variants \cite[e.g.][]{Baluev-08}, but it is based on a different principle. The $l_1$  periodogram can analyze the radial velocity without the need to estimate the frequency iteratively, using the theory of compressed sensing adapted for handling correlated noise. Instead of fitting a sinusoidal function at each frequency as the Lomb-Scargle-type periodograms, the $l_1$ periodograms searches directly for a representation of the input signal as a sum of sinusoids. As a result, all the frequencies of the signal are searched simultaneously. This relies on the so-called "sparse recovery" tools, which are designed to find a representation of an input signal as a linear combination of sine functions.\\

The $l_1$ periodogram outputs a figure which has a similar aspect as a standard  Lomb-Scargle-type periodogram, but with fewer peaks due to aliasing. We use the $l_1$ periodogram of the HARPS data on a grid of frequencies between 0.01 and 0.3 cycles per day. In order to measure the significance of a possible signal the False Alarm Probability (FAP) is calculated, which encodes the probability of measuring a peak of a given height conditioned on the assumption that the data consists of Gaussian noise with no periodic component.  Typically a signal at the 1$\%$ FAP level is considered suggestive, and 0.1 $\%$ FAP level statistically significant. Figure 5 shows two $l_1$ periodograms of the HARPS RVs. The top panel corresponds to the undetrended RVs. In this case we detect three significant signals at 4.11d (FAP $1.10^{-4}$\%), 7.98 days (FAP $7.10^{-6}$\%) and 26.6 days (FAP $3.10^{-5}$\%). Although the coefficient amplitude of the second and third peak are relatively close, their FAP can differ significantly  since it is estimated in a sequential manner, i.e. the highest peak is computed against zero, the second peak against the first and the third peak against the second. Figure \ref{fig:fwhm} clearly shows the effects of activity in the FWHM  at the end of the time series. It suggests that the star was more active at that moment, probably due to one spot crossing event during one single rotational period, standing for only 20 days (from BJD 58830 to 58850). However, we find that signal at 26.6 days disappears when we detrend the RVs using GP modelling with \textit{SPLEAF} \citep[e.g.][]{spleaf} (see bottom panel of Figure 5). The GP is trained on the $log R^{\prime}_\mathrm{HK}$ indicator and then it is used to detrend the RVs using a linear fit. This method gives two significant peaks at at 4.11d with FAP $6.10^{-5}$\% and 8.00 days with FAP $4.10^{-4}$\%. 

Another way to detrend the stellar activity is to fit the correlations between activity indicators as the full width at half maximum (FWHM), the bisector or the S-index, and subtract them in order to correct for any long-term effects resulting from a magnetic cycle. We do so dividing the data into different time regimes selected in order to maximize the correlation of the  RV data with the activity indicators in each bin. This approach also identifies two peaks at 4.11d with FAP $7.10^{-8}$\% and 8.01 days with FAP $3.10^{-4}$\%. Finally, we also find the same two significant signals when we do not use the data TBJD 58730 to  TBJD 58750, which corresponds to period in which the star was more active. The signal at 26.6 days only appears when using the full undetrended RV sample which is clearly affected by stellar activity and it is within 1.6-sigma of the rotation period derived from the chromospheric activity indicator. We therefore conclude that only the peaks at $\sim$4.11 days and ~8 days are caused by planets. In addition, despite the ~8 day signal is nearly double the $\sim$4.11 day, in all the cases presented we find the ~8 day signal after removing the $\sim$4.11 day signal. We also evaluate the evidence of the second planet by comparing the difference in Bayesian Information Criterion ($\Delta{\rm BIC}$) between the one-eccentric-planet and the two-eccentric-planets models. We get a value of $\Delta{\rm BIC}$=11, which suggests a strong evidence.  \\

 


\begin{table*}
\begin{minipage}{12cm}
\caption{TOI-1062 parameters from \juliet: median and 68\% confidence interval.}
\begin{tabular}{lccc}\hline

\smallskip\\\multicolumn{2}{l}{\underline{Instrumental Parameters:}}&\smallskip\\

~~~~$q_{1,TESS}$\dotfill &Quadratic limb-darkening parametrization\dotfill &$0.842^{+0.037}_{-0.035}$&\\
~~~~$q_{2,TESS}$\dotfill &Quadratic limb-darkening parametrization\dotfill &$0.114^{+0.113}_{-0.069}$&\\
~~~~$m_{flux,TESS}$\dotfill &Offset relative flux\dotfill &$-0.0000215^{+0.0000016}_{-0.0000018}$\\
~~~~$\sigma _{TESS}$\dotfill &Jitter (ppm)\dotfill &$118.56^{+5.24}_{-5.01}$\\
~~~~$\sigma _{HARPS}$\dotfill &Jitter (m/s)\dotfill& $1.127^{+0.149}_{-0.088}$\\

\smallskip\\\multicolumn{2}{l}{\underline{Parameters of the GP with quasi-periodic kernel:}}&\smallskip\\

~~~~$GP_{\sigma}$\dotfill &Amplitude (m/s)\dotfill &$4.8^{+0.83}_{-0.58}$&\\
~~~~$GP_{\rho}$\dotfill &Length-scale of the Matern part (days)\dotfill &$4.70 ^{+1.24}_{-0.87}$&\\
~~~~$GP_T$\dotfill &Length-scale of exponential part (days)\dotfill &$31.04^{+10.87}_{-10.12}$\\


\smallskip\\\multicolumn{2}{l}{\underline{Planetary Parameters:}}&b&c\smallskip\\

~~~~$P$ \dotfill &Period (days) \dotfill &$4.11412^{+0.0014}_{-0.0015}$&$7.978^{+0.021}_{-0.023}$\\
~~~~$T_0$ \dotfill &Time of transit center$^*$ ($BJD_{TBD}$) \dotfill &$2459082.58^{+0.09}_{-0.08}$& $2459087.58^{+1.86}_{-1.64}$\\
~~~~$K$\dotfill & Radial velocity semi-amplitude (m/s) \dotfill &$4.28^{+0.32}_{-0.34}$&$3.87^{+0.52}_{-0.53}$\\
~~~~$\sqrt{e} sin(\omega )$\dotfill &Parametrization for $e$ and $\omega$\dotfill &$0.15^{+0.09}_{-0.08}$& $-0.02^{+0.06}_{-0.07}$  \\
~~~~$\sqrt{e} cos(\omega )$\dotfill &Parametrization for $e$ and $\omega$\dotfill &$-0.06^{+0.07}_{-0.06}$& $-0.09^{+0.11}_{-0.10}$\\
~~~~$r_1$\dotfill &Parametrization for p and b \dotfill &$0.909^{+0.005}_{-0.007}$&-\\
~~~~$r_2$\dotfill &Parametrization for p and b \dotfill &$0.0233^{+0.0003}_{-0.0004}$& - \\

\smallskip\\\multicolumn{2}{l}{\underline{Derived transit and RV parameters:}}&b&c\smallskip\\

~~~~$e$\dotfill &Eccentricity or the orbit  \dotfill &$0.179^{+0.080}_{-0.061}$& $0.137^{+0.088}_{-0.066}$ \\
~~~~$\omega$\dotfill & Argument of periastron (deg)  \dotfill &$110^{+29}_{-20}$& $151^{+22}_{-64}$ \\
~~~~$i$\dotfill & Inclination (deg)  \dotfill &$85.908^{+0.061}_{-0.043}$& - \\
~~~~$p=R_p/R_{\star}$\dotfill &Planet-to-star radius ratio \dotfill &$0.0234^{+0.0004}_{-0.0004}$& - \\
~~~~$b$\dotfill &Impact parameter of the orbit \dotfill &$0.862^{+0.007}_{-0.009}$& - \\

\smallskip\\\multicolumn{2}{l}{\underline{Derived physical parameters:}}&b&c\smallskip\\

~~~~$M_p$\dotfill & Planetary mass ($M_{\oplus}$)  \dotfill &$10.17^{+0.82}_{-0.86}$& - \\
~~~~$R_p$\dotfill & Planetary radius ($R_{\oplus}$)  \dotfill &$2.264^{+0.095}_{-0.091}$& - \\
~~~~$\rho _p$\dotfill & Planetary density ($g/cm^3$)  \dotfill &$4.86^{+0.85}_{-0.74}$& - \\
~~~~$a _p$\dotfill & Semi-mayor axis ($AU$)  \dotfill &$0.052^{+0.023}_{-0.025}$& $0.080^{+0.013}_{-0.012}$ \\
~~~~$S$\dotfill & Insolation ($S_{\oplus}$)  \dotfill &$232^{+11}_{-10}$& $95^{+5}_{-4}$ \\
~~~~$T_{eq}$\dotfill & Equilibrium Temperature ($K$)  \dotfill &$1077^{+10}_{-9}$& $859^{+9}_{-8}$ \\
~~~~$M_P\sin i$\dotfill &Minimum planetary mass ($M_{\oplus}$)\dotfill &$10.17^{+0.87}_{-0.85}$& $9.35^{+1.23}_{-1.23}$ \\

\\

\hline
\end{tabular}
   \vspace{1ex}

     {\raggedright * For TOI 1062c $T_0$ corresponds to the time when the planet would have transited if it did. \par}
\end{minipage}
\end{table*}


\subsection{Global fit with \juliet}
\label{sec:exofast}

We determine the planetary parameters using the publicly available software \juliet \cite[e.g.,][]{juliet}. \juliet allows us to jointly fit the TESS and LCOGT photometry and the HARPS radial velocities with GP. The software is built on several publicly available tools for the modeling of the photometric data  \citep[using the \batman package,][]{kreidberg2015}, the radial velocities \citep[using the \radvel package,][]{fulton2018}, and also to incorporate GP \citep[via the \celerite package,][]{foreman2017}. The parameter space is explored using nested sampling, with the \multinest algorithm \citep[e.g.,][]{Feroz08} in its Python implementation, \pymultinest  \cite[e.g.,][]{buchner-14}. In addition, \juliet computes the Bayesian evidence using \dynesty \cite[e.g.,][]{speagle2020}, a Python package to estimate Bayesian posteriors and evidences using Dynamic Nested Sampling. In short, nested sampling algorithms work as follows. The algorithm samples some number of live points randomly from the prior distribution, and the likelihood is evaluated at each of these points. At each iteration the point with the lowest likelihood is replaced by a new point sampled, keeping the number of live points constant. This process is continued until Bayesian evidence reaches some specified value. The number of live points used has to be large enough to adequately sample the parameter space.\\

The transit model fits the stellar density $\rho_{\star}$ together with the planetary and jitter parameters. For the stellar density we use the value obtained in Section 3, and the priors of the orbital parameters of the inner planet are taken from ExoFOP. We use the quadratic limb darkening coefficients $(q_1,q_2)$ introduced by \cite{kipping2013} for the photometric data, since it was shown to be appropriate for space-based missions \citep{espinoza2015}. In addition, instead of fitting the planet-to-star radius ratio and the impact parameter of the orbit,  use the parametrization introduced in \cite{espinoza2018} and fit the parameters $r_1$ and $r_2$ to make sure the full exploration of physically plausible values in the (p,b) space. We parametrize the eccentricity and the argument of periastron with $\sqrt{e}\sin{\omega}$ and $\sqrt{e}\cos{\omega}$, always ensuring that $e\leq1$. We use the \celerite aproximate Matern multiplied by exponential kernel to account for the stellar activity. The fit obtained with this kernel is nearly identical to the one obtained with other kernels as the quasi-periodic or the exp-sine-squared kernel.  \\

\begin{figure}[t!]
\begin{center}
\includegraphics[scale=0.55]{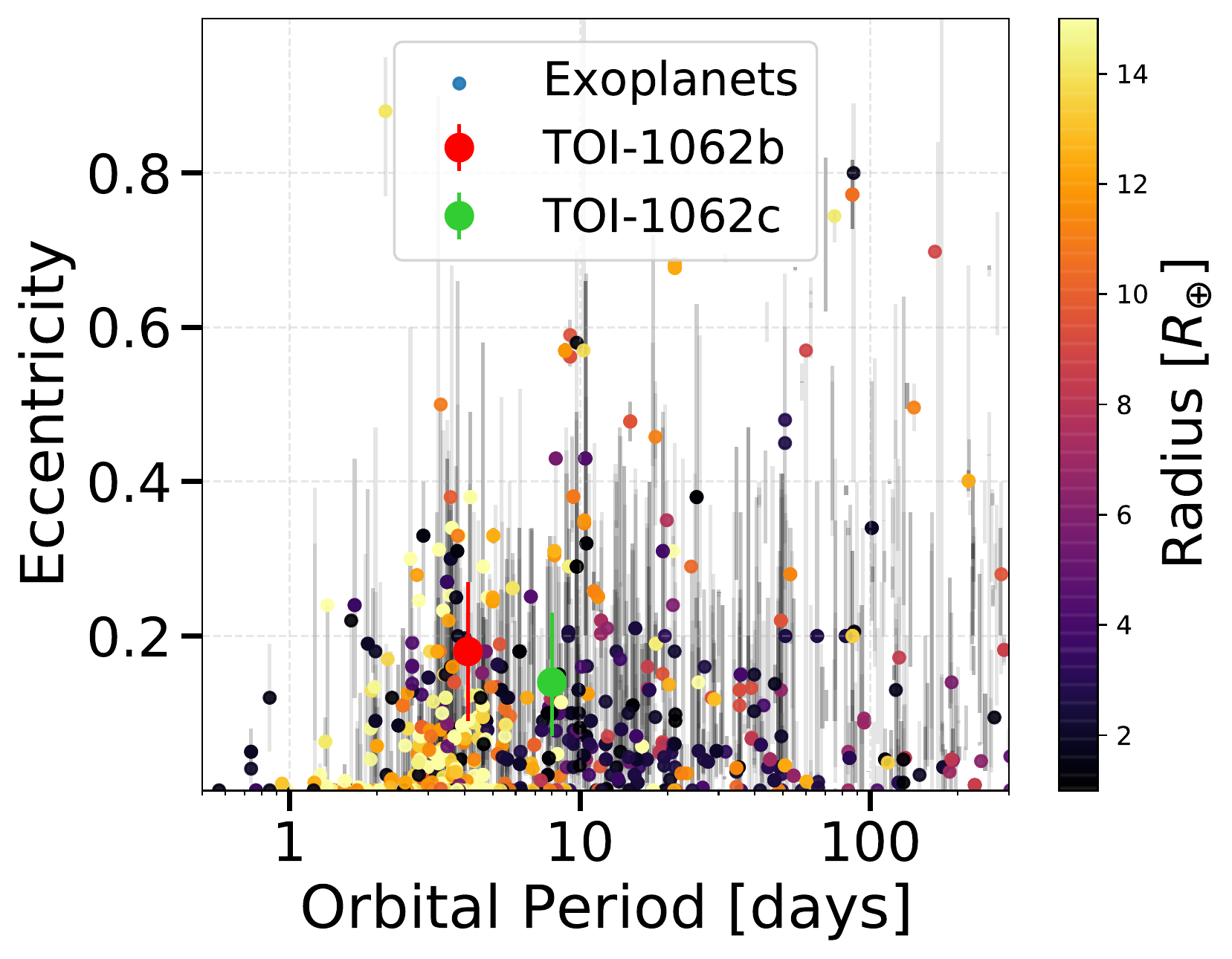}
\caption{Orbital eccentricity against orbital period for observed exoplanets from the NASA Exoplanet Archive. The color represents the planetary radius. }  
\label{Eccentricity}
\end{center}
\end{figure} 

We fit both planets with \juliet and find no hint of transit for TOI\,1062c, as shown in Figure \ref{fig:light_curve}. The final median planetary parameters determined by the \juliet fit are listed in Table 2. We find that TOI\,1062b has an orbital period of 4.114 days, a radius of $2.264^{+0.095}_{-0.091}$ $R_{\oplus}$, a mass of $10.17^{+0.83}_{-0.81}$ $M_{\oplus}$ and a density of $4.86^{+0.85}_{-0.74}$ g cm$^{-3}$, slightly below the Earth's mean density. Nevertheless, as discussed in detail in Section 5, its composition and internal structure are unlikely to be similar to that of the Earth. We find an eccentricity of $0.18^{+0.08}_{-0.07}$ for TOI-1062b and $0.14^{+0.09}_{-0.07}$ for TOI-1062c. Figure 8 shows the eccentricity against the orbital period for the population of observed exoplanets from the NASA Exoplanet Archive, and we see that TOI-1062b is slightly higher than usual for its orbital period. However, it may not be significant due to the bias towards larger eccentricities for nearly circular planets \cite[][]{Lucy-71}. TOI\,1062\,c instead is found to have an orbital period of 7.988 $\pm$ 0.04 days and a minimum mass of 9.35 $\pm$ 1.23 M$_{\oplus}$. TOI\,1062\,c is in near the 2:1 motion resonance with its inner companion.


\section{Discussion}

\subsection{Internal structure}

In order to characterize the internal structure of TOI\,1062~b, we model its interior considering a pure-iron core, a silicate mantle, a pure-water layer, and a H-He atmosphere. The equations of state (EOSs) used for the iron core are taken from \cite{Hakim2018},  the EOS of the silicate-mantle is calculated with PERPLE\_X from \cite{Connolly09} using the thermodynamic data of \cite{stixrude_thermodynamics_2011} and assuming Earth-like abundances, and the EOS for the H-He envelope are from \cite{Chabrier2019} assuming a proto-solar composition.  For the pure-water layer, we use the AQUA EOS from \cite{Haldemann2020}. We assume an envelope luminosity of L=$10^{22.52}$ erg\,s$^{-1}$ (equal to Neptune's luminosity). The thickness of the planetary layers were set by defining their masses and solving the structure equations. To obtain the transit radius, we follow \cite{Guillot2010} and evaluate the location where the chord optical depth $\tau _{ch}$ is $2/3$. We do not use stellar abundances as an additional constraint since it is not clear whether they are a good proxy for the planetary bulk abundances \cite[][]{Wang-19,Plotnykov} and it has been shown that the stellar abundances are not always useful for constraining the internal composition \cite[][]{otegi-20-2}. \\

  \begin{figure*}[h]
  \centering
  \begin{tabular}{@{}cc@{}}
    \includegraphics[scale=0.5]{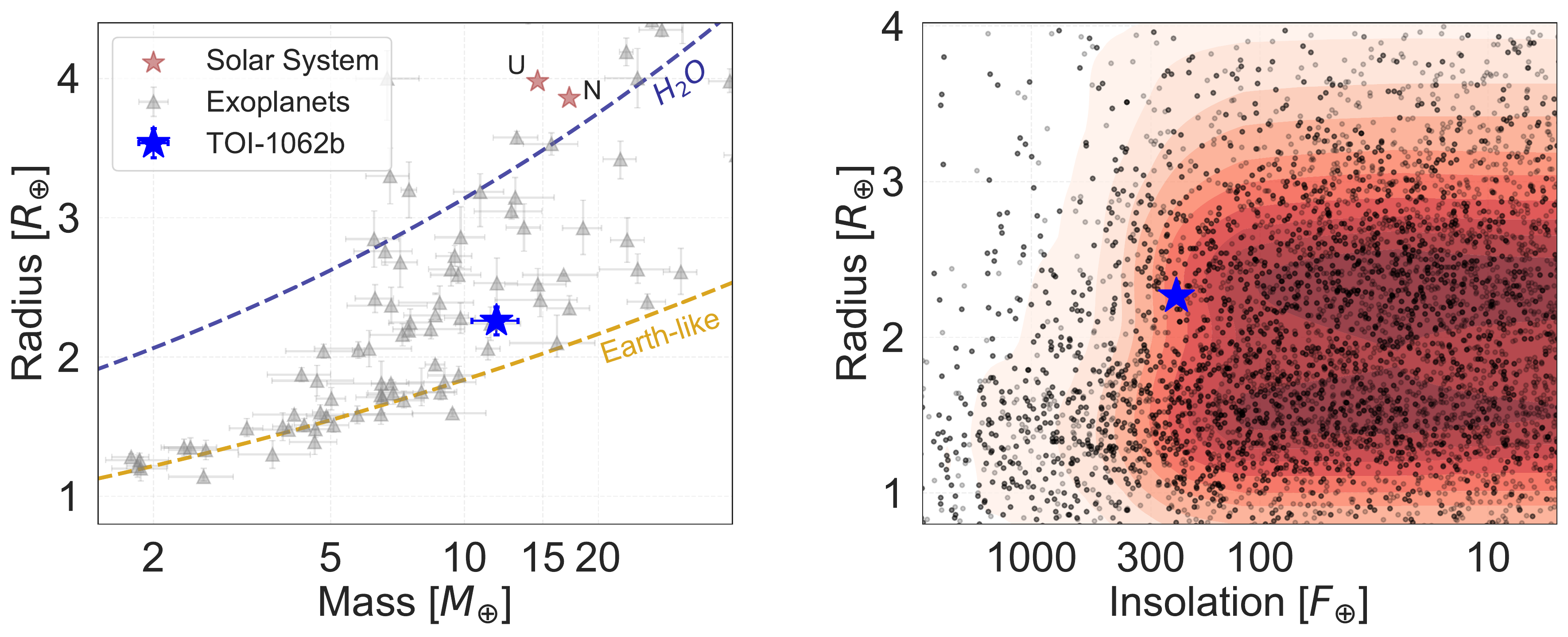}
    \end{tabular}
    \caption{ Left: Mass-radius diagram of exoplanets with accurate mass and radius determination from \cite{otegi2020}. Also shown are the composition lines for Earth-like planets and pure water subjected to a stellar radiation of $F/F_{\oplus}$= 230 (similar to the one of TOI\,1062~b). Right: Radius against insolation flux for known exoplanets from NASA Exoplanet Archive. The colormap indicates the  point density. TOI\,1062\,$b$ is represented by a blue star.  \label{fig:mr} }

\end{figure*}

Figure \ref{fig:mr} shows M-R curves tracing the compositions of Earth-like (with a CMF=0.33) and pure-water subjected to a stellar radiation of $F/F_{\oplus}$= 230 (similar to that of TOI\,1062~b). For reference, we also show exoplanets with accurate and reliable mass and radius determinations  \cite[][accessible on the Data \& Analysis Center for Exoplanet  DACE\footnote{https://dace.unige.ch/exoplanets/}]{otegi2020}. TOI\,1062~b sits above the Earth-like curve and  below the pure-water curve, suggesting that it contains a small amount of volatile materials of less than 1\% in mass.  Figure ~\ref{fig:mr} also displays the insolation flux relative to Earth against radii for the known exoplanets extracted from the NASA Exoplanet Archive, which shows the separate populations of super-Earths and mini-Neptunes. We see that TOI\,1062~$b$ sits in the mini-Neptune regime, close to the radius valley. In \cite{otegi2020} we identified two distinct populations of volatile-rich and "rocky" (referring to exoplanets expected to have  low amounts of volatiles) separated by the water-line. Using these results, we see that TOI\,1062b lies in the "rocky" exoplanet regime defined in \cite{otegi2020} even if its radius is above the radius valley, suggesting that it is mostly composed of refractory materials by mass.  \\

\begin{figure*}[h]
\centering
  \begin{tabular}{@{}cc@{}}
    \includegraphics[scale=0.47]{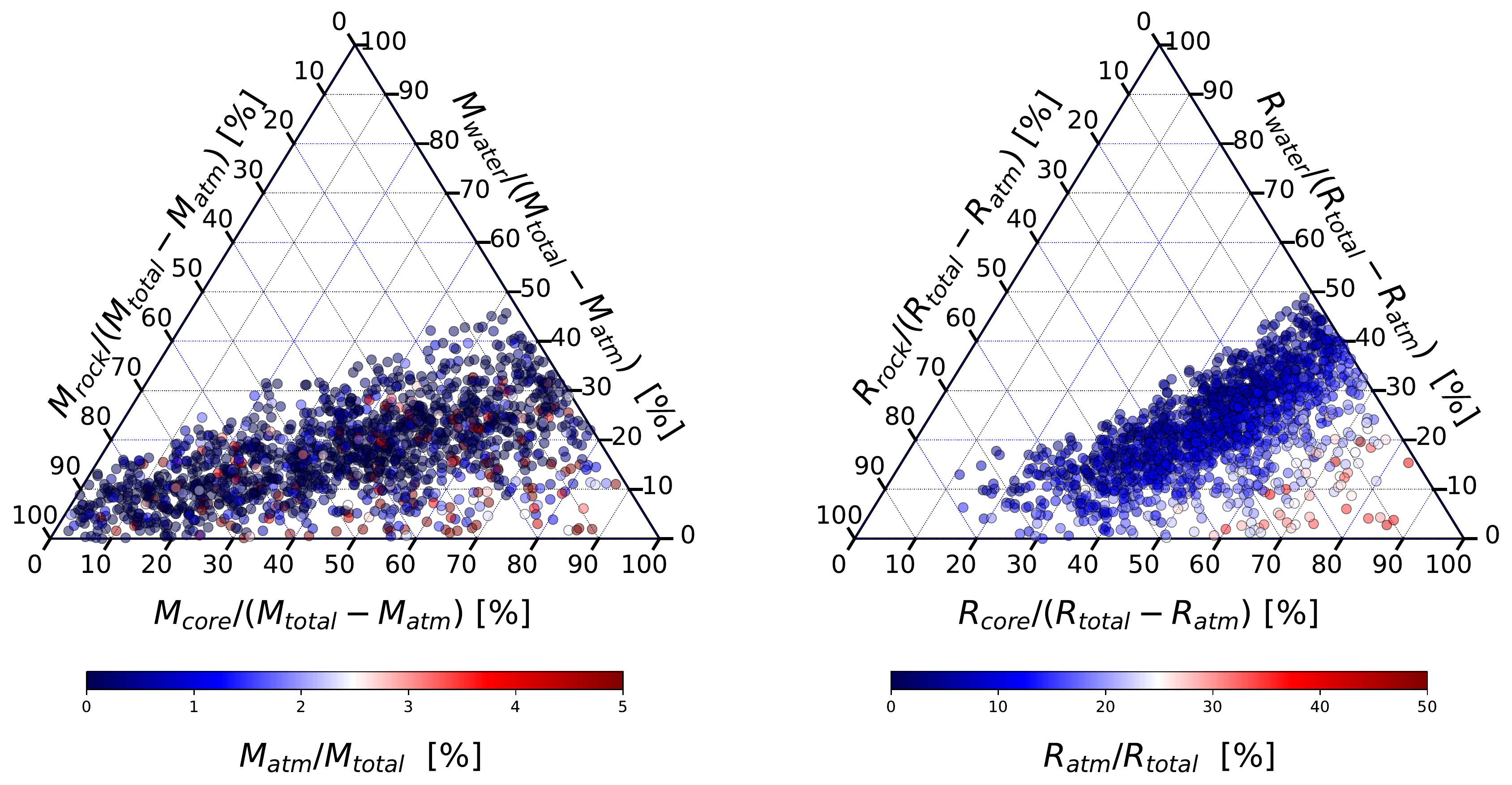}
  \end{tabular}
  \caption{ Ternary diagram of the inferred internal composition of TOI\,1062~b. We show the parameters space covered by the posterior distributions in the 4-layer model in mass (left) and radius (right). \label{fig:interior}}
\end{figure*}

We use a generalized Bayesian inference analysis using a Nested Sampling scheme \citep{buchner-14} to quantify the degeneracy between various interior parameters and produce posterior probability distributions. Fig.~\ref{fig:interior} shows ternary diagrams of the inferred composition of TOI\,1062~b. The ternary diagram shows the degeneracy associated with the determination of the composition of exoplanets with measured mass and radius. 
We find a median H-He mass fraction of 0.1\%, which corresponds to a lower-bound since enriched H-He atmospheres are more compressed and, therefore, increase the planetary H-He mass fraction. Indeed, formation models suggest that sub-Neptunes are likely formed by envelope enrichment \cite[][]{Venturini2017}. We also find that TOI\,1062~b is expected to have a very significant iron core and silicate mantle, accounting for nearly 40\% of the planetary mass and thicknesses of 1$R_{\oplus}$ and 0.5$R_{\oplus}$ respectively. The water layer has an estimated relative mass fraction of 15\%. Nevertheless, the degeneracy between the core, silicate mantle, and water layer in this M-R regime is particularly high \cite[][]{otegi-20-2}, and it does not allow accurate  estimates of the masses of these constituents.  
Interior models cannot distinguish between water and H-He as the source of low density material, so we also run a 3-layer model without the H$_{2}$O envelope. Table \ref{tab:table_comp} lists the inferred mass fractions of the core, mantle,water-layer and H-He envelope for both the 4-layer model and the water-free one. In this case we find that the planet is 0.35\% H-He, 26\% iron and 73\% rock by mass, setting maximum limits for the atmospheric and rock mass since any water added would decrease these mass fractions.

\begin{table}
\centering
\caption{Inferred interior structure properties of TOI\,1062b.}
\label{tab:table_comp}
\begin{tabular}{lcc}
\hline
\textbf{Constituent} & \textbf{4-layer [\%] }& \textbf{Without H$_{2}$O [\%] }\\

\hline
$M_{\rm core}/M_{\rm total}$ & $44 ^{+26} _{-24} $ &  $26 ^{+26} _{-11}$\\
$M_{\rm mantle}/M_{\rm total}$ & $ 40 ^{+27} _{-16} $ &  $73 ^{+11} _{-14}$\\
$M_{\rm water}/M_{\rm total}$ &  $15 ^{+9} _{-6}$ & -\\
$M_{\rm H-He}/M_{\rm total}$ & $0.11 ^{+0.06} _{-0.04}$ & $0.35  ^{+0.12} _{-0.08}$\\
\hline

\end{tabular}

\end{table}

\subsection{Dynamical analysis}
As a general note of this section, the studies presented here were done in the hypothesis of a co-planar planetary system, i.e. the orbits of planets b and c evolve in the same plane. This is consistent with the observations, since TOI\,1062c would not transit if it had the same orbital inclination as TOI\,1062b. 
\subsubsection{Resonance}

As can be seen from Table 2, the period ratio of the planets is $P_c/P_b = 1.941$. The planet pair thus lies close to the 2:1 mean-motion resonance (MMR). How close is the system to the MMR? We explore the structure of the parameter space of TOI-1062 near that resonance, in order to investigate its dynamical state. \\ 

\begin{figure}[h]
\begin{center}
\includegraphics[width=\columnwidth]{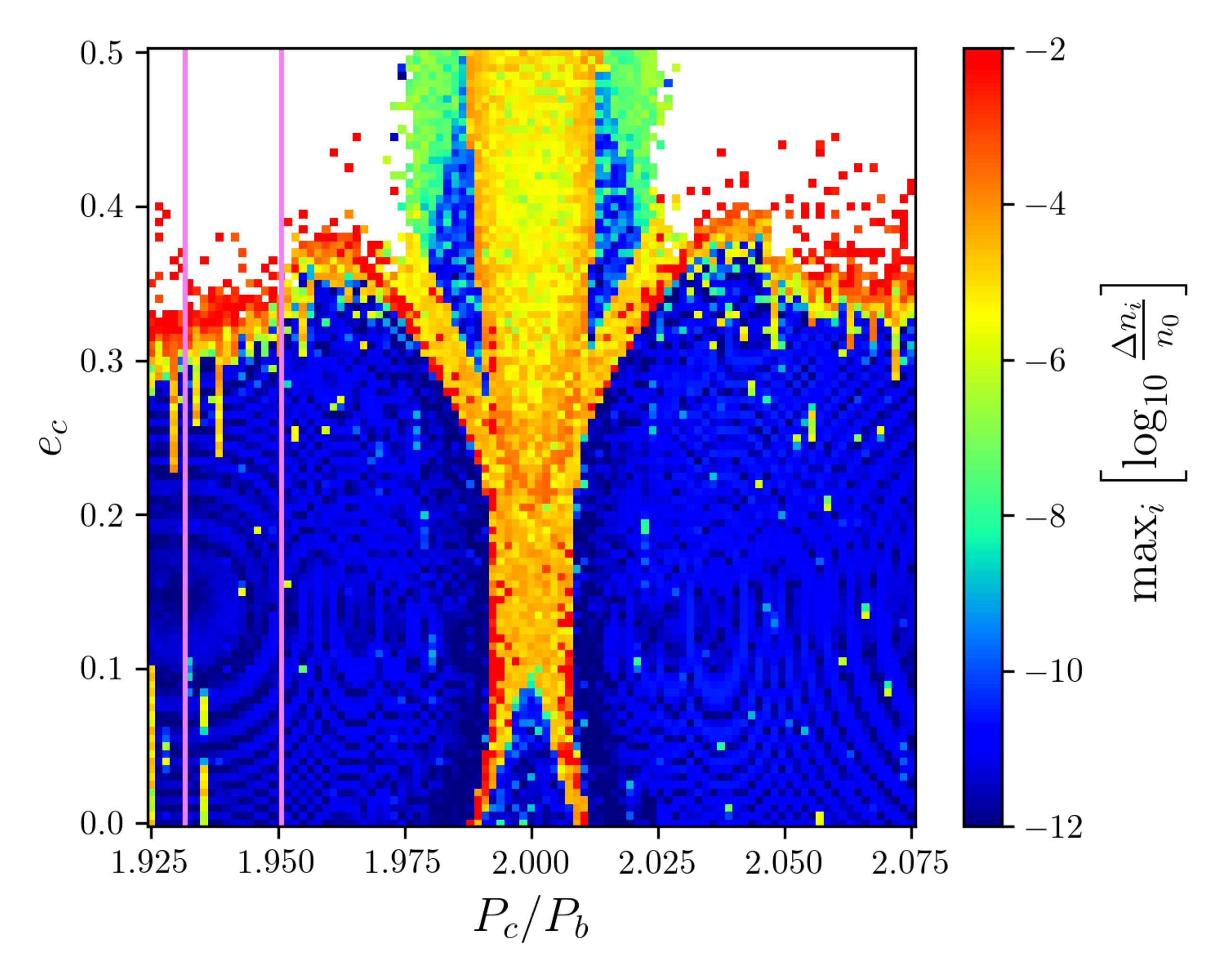} 
\caption{Chaoticity map around the solution for the 2-planet model (Table 2). The eccentricity of the outer planet $e_c$ and the period ratio $P_c/P_b$ are explored on a grid of 101x101 different configurations. A color is assigned to each box after a short numerical integration, according to its level of chaos using the NAFF indicator (see text).}  
\label{StMap}
\end{center}
\end{figure}

We therefore  design a two dimensional section of the parameter space defined by the period ratio on one axis and the eccentricity of the outer planet $e_c$ on the other. Our section has a resolution of 101x101, i.e., we explore the dynamics of 10201 initial configurations defined by a unique set of ($P_c/P_b$,$e_c$). All the other orbital parameters were initially fixed at the best values reported in Table 2. We numerically compute the future evolution of each configuration over 20 kyr. This is performed with the adaptive time-step high order N-body integrator IAS15, which is available from the python package REBOUND\citep{Rein2012, Rein2015}. 
The perturbative effect of general relativity described in \citet{Anderson1975} was included via the library REBOUNDx \citep{Tamayo2019}. From these numerical simulations, the level of chaos of each configuration was evaluated with the NAFF fast chaos indicator \citep{Laskar1992, Laskar1993}. The result is presented in Fig. \ref{StMap}. \\ 

The NAFF computes precisely, for each planet, the average mean motion $n=\frac{2\pi}{P}$ over the two halves of the integration and compares these two estimations. Due to the secular constancy of the semi-major axis of regular orbits, this difference should be small in non-chaotic orbits. The higher is the drift in the average mean-motion, the more chaotic is the orbit. In this work, we took as the NAFF of the system the maximal value of this drift over the planetary orbits, in logarithmic scale: $NAFF ~ = ~ \max_i \left[log_{10} \frac{\Delta n_i}{n_{0,i}}\right]$, where the subscript $i$ refers to the planet $b$ or $c$, $\Delta n_i$ is the drift of the average mean motion over the two halves of the integration, and $n_{0,i}$ is the initial mean motion of planet $i$. The color code in Fig.~\ref{StMap} depicts the NAFF defined here above: the bluer, the more regular is the system. We also distinguish the strongly chaotic configurations from the ones that did not finish the integration because of either a close-encounter or an escape of a body (white boxes). 
We finally explain our choice of 20 kyr for the total integration time. Over this time span, the planetary orbits in the TOI-1062 system are expected to cover several secular cycles during which the semi-major axes oscillate. Covering several of these cycles allows to properly average the secular variations and isolate the chaotic diffusion. We verified numerically that several secular cycles are made up over an integration. \\ 


In this map, the 2:1 mean motion resonance appears clearly as the orange band in the middle of the plot. For the current system's parameters and the resolution of our chaoticity map, this resonance seems therefore chaotic. It is important to stress that this picture is highly dependent on the system's parameters. For instance in that case, the arguments of periastron $\omega_b$ and $\omega_c$ are close to the alignement. The opposite configuration where the orbits are anti-aligned would show a drastically different picture, with a different strength and apparent stability of the 2:1 MMR. \\ 
The two vertical lines depict the $1\sigma$ window of $P_c$. With the currently estimated orbital parameters and planet masses, the TOI\,1062 system certainly lies outside of the 2:1 MMR. Despite the influence that a revision of the parameters may have on the resonance, it seems very unlikely that this conclusion changes. \\ 

Finally, we stress that the uncertainty on the eccentricity of the inner planet $e_b$ is somewhat large. Modifying this parameter will directly impact the strength of the 2:1 MMR, as the resonance width is expected to increase with the eccentricity. The same note applies for the planetary masses. \\ 


\subsubsection{Stability constraints}
While the chaoticity map presented above (Fig. \ref{StMap}) is informative about the maximal eccentricity of the outer planet allowed by orbital stability, this estimate should be considered with caution. Indeed, this map explores the parameter space on two dimensions, without taking into account potential correlations with the other parameters. As already discussed, if the best fit estimate of the other parameters changes, the entire map might change as well. 
In order to give proper stability constraints on the orbital parameters and masses of the planets, we also perform a global stability analysis from the posterior distribution of the joint photometry + RV analysis. The only fixed parameter is the inclination of the outer planet, taken equal to the inclination of the inner planet (i.e. we explored the co-planar case, which is compatible with the outer planet non transiting). We selected a sample of 10000 solutions from that posterior. Each of these is numerically integrated over 20 kyr using the same integrator set-up described in the previous section. The level of chaos is again computed with the NAFF indicator. Over the 10000 configurations, 8455 did survive the entire 20 kyr simulation. The others were thus unstable (escape or close encounter of two bodies). Imposing a more strict stability threshold by further removing all solutions with $NAFF > -5$ leaves us with a stable posterior of 7224 configurations. \\ 

In any case, no stringent constraint can be added on the orbital parameters and masses of the planets. With the sorting in NAFF as defined above, only a slight cut in the eccentricities is observed. The new median estimates and $1\sigma$ confidence intervals are $e_b = 0.162$ [$0.098$,$0.228$] and $e_c = 0.129$ [$0.065$,$0.205$]. In particular, the global stability analysis does not allow us to further constrain the planetary masses (again, assuming co-planar orbits).


\subsection{Atmospheric characterization}

TOI\,1062b poses an interesting target for atmospheric characterisation given its equilibrium temperature is $\sim$1000~K. We expect a high metallicity atmosphere for a planet with such a small radius and mass, as a result of the strong stellar irradiation received which would result in a significant loss of the H$_2$/He envelope. However, such a target is an excellent laboratory to study carbon chemistry at high temperature given its high expected metallicity. Beyond metallicities of $\gtrsim300\times$ solar, the primary carbon bearing species in the atmosphere transitions from being CH$_4$/CO to CO$_2$ in thermochemical equilibrium \citep[e.g.][]{moses2013_gj436}. This planet is thus an ideal target to explore the CO, CH$_4$, CO$_2$ chemical stability boundary which occurs at $\sim$800~K in photospheric conditions at such high metallicities. The mean molecular weight for such an atmosphere at $\sim300\times$ metallicity is expected to be $\sim$4~g/mol, and hence significant features are still present in transmission spectra compared to a solar metallicity atmosphere (at $\sim2.35$~g/mol). Using the transmission spectroscopy metric (TSM) defined in \citet{kempton2018}, we determine the that TOI\,1062b has a TSM value of 31.8 assuming a mean molecular mass of 2.3~g/mol and 18.3 assuming 4 g/mol. The lower TSM value for the higher mean molecular weight is due to the reduced atmospheric scale height of a heavier atmosphere. This value is comparable to targets such as Trappist-1f and several simulated TESS targets from \citet{sullivan2015}. 

On the other hand, TOI\,1062b may have completely lost all of its H$_2$/He envelope which would result in an ultra-high metallicity secondary atmosphere. At metallicities $\gtrsim3000\times$ solar, the mean molecular weight of the heavier species such as CO$_2$ and H$_2$O now completely dominate due to the lack of significant H$_2$ or He. Therefore, the mean molecular weight is likely to be $>18$~g/mol, and thus the scale height of spectral features in the transmission spectra are significantly reduced by a factor of $\gtrsim7\times$ over a solar metallicity atmosphere. This reduces the TSM value for such an atmosphere to $\lesssim$5, making constraints on the abundances difficult. However, observations of the terminator may still be able to place a lower limit on the molecular weight and thus the metallicity. These would also provide a direct contrast to planets such as 55 Cancri e which also indicate a high mean molecular weight atmosphere \citep{abhinav2020}. In addition, secondary eclipse spectroscopy for such a cool target will be challenging, but has been achieved for GJ436b \citep{stevenson2010}. However, for the case of TOI\,1062b, the host star is a G-type and thus will result in a very weak planet/star flux contrast in emission spectroscopy.

\section{Conclusions}

We present the discovery of two new planets from the TESS mission in the TOI\,1062 system. The analysis is based on 2-min cadence TESS observations from 4 sectors, ground-based follow-up from LCOGT and RV data from the HARPS spectrograph. High-resolution imaging from Zorro speckle imager rules out  the presence of nearby companions and potential nearby eclipsing binaries. We find that the host star is rotating with a period of 21.9 days, as well as the existence of a stellar spot whose life-time is similar to the stellar rotational period. \\

TOI\,1062b is expected to be a mini-Neptune with a period of 4.11 days, radius of 2.26$\pm$0.1 $R_{\oplus}$, and mass of 10.2$\pm$0.8 $M_{\oplus}$. Internal structure models indicate that TOI\,1062b is expected to be composed of significant iron core and silicate mantle, accounting for nearly 40\% of the planetary mass each, and that it is expected to have a small volatile envelope of 0.35\% of the mass at maximum.
TOI\,1062c, which is not transiting, is found in the HARPS RV data. We find that its minimum mass is inferred to be 9.35$\pm$1.23 $M_{\oplus}$, and that it is close to a 2:1 motion resonance with its inner companion, with a period of 8.12 days. 
The position of TOI\,1062b with respect to the radius valley and its high equilibrium temperature of ~1000K make it a interesting candidate for atmospheric characterization. The strong stellar irradiation may result in a significant loss of the H-He envelope leaving a high metallicity atmosphere that can be an excellent laboratory to study carbon chemistry at high temperature.

\begin{acknowledgements}
This work has been in particular carried out in the frame of the National Centre 
for Competence in Research ‘PlanetS’ supported by SNSF. D.J.A. acknowledges support from the STFC via an Ernest Rutherford Fellowship (ST/R00384X/1). This work was supported by FCT - Funda\c{c}\~ao para a Ci\^encia e a Tecnologia through national funds and by FEDER through COMPETE2020 - Programa Operacional Competitividade e Internacionaliza\c{c}\~ao by these grants: UID/FIS/04434/2019; UIDB/04434/2020; UIDP/04434/2020; PTDC/FIS-AST/32113/2017 \& POCI-01-0145-FEDER-032113; PTDC/FIS-AST/28953/2017 \& POCI-01-0145-FEDER-028953.  V.A. and E.D.M acknowledge the support from FCT through Investigador FCT contracts nr.  IF/00650/2015/CP1273/CT0001, IF/00849/2015/CP1273/CT0003, respectively.
C.D. acknowledges support from the Swiss National Science Foundation under grant PZ00P2\_174028. Siddharth Gandhi acknowledges support from the UK Science and Technology Facilities Council (STFC) research grant ST/S000631/1. Resources supporting this work were provided by the NASA High-End Computing (HEC) Program through the NASA Advanced Supercomputing (NAS) Division at Ames Research Center for the production of the SPOC data products. We acknowledge the use of public TESS Alert data from pipelines at the TESS Science Office and at the TESS Science Processing Operations Center. This research has been partly funded by the Spanish State Research Agency (AEI) Projects No.ESP2017-87676-C5-1-R and No. MDM-2017-0737 Unidad de Excelencia "Mar\'ia de Maeztu"- Centro de Astrobiolog\'ia (INTA-CSIC). S.G.S acknowledges the support from FCT through Investigador FCT contract nr. CEECIND/00826/2018 and POPH/FSE (EC). H.P.O. acknowledges that this work has been carried out within the framework of the NCCR PlanetS supported by the Swiss National Science Foundation. S.H. acknowledge support by the fellowships PD/BD/128119/2016 funded by FCT (Portugal). X.D is grateful to The Branco Weiss Fellowship--Society in Science for its financial support.  This project has received funding from the European Research Council (ERC) under the European Union's Horizon 2020 research and innovation programme (grant agreement No 851555/SCORE. A.O acknowledges support from an STFC studentship. S.H acknowledges CNES funding through the grant 837319.This work was supported by FCT through national funds (PTDC/FIS-AST/28953/2017) and by FEDER - Fundo Europeu de Desenvolvimento Regional through COMPETE2020 - Programa Operacional Competitividade e Internacionaliza\c{c}\~ao (POCI-01-0145-FEDER-028953) and through national funds (PIDDAC) by the grant UID/FIS/04434/2019. Funding for the TESS mission is provided by NASA's Science Mission directorate. 

This work makes use of observations from the LCOGT network.
\end{acknowledgements}

\bibliography{bibliography}

\appendix
\section{HARPS spectroscopy}

\begin{table}
\caption{HARPS spectroscopy obtained between 31st August 2019 and 11th November 2019.}
\label{tab:spec1}
\scriptsize
\begin{tabular}{lcccccc}
\hline
\hline
Time & RV & $\sigma_{\rm RV}$ & $S_{\rm MW}$ & $\sigma_{S}$ & FWHM & $\sigma_{\rm FWHM}$ \\
$[\rm{BJD}-2400000]$ & \multicolumn{2}{c}{[m/s]} & \multicolumn{2}{c}{--} & \multicolumn{2}{c}{[m/s]} \\
\hline
\hline
$58726.77$ & $-4.63$ & $1.19$ & $0.2308$ & $0.0035$ & $6521.73$ & $9.7$ \\
$58726.83$ & $-3.58$ & $1.39$ & $0.2205$ & $0.0051$ & $6516.21$ & $15.2$ \\
$58727.78$ & $-4.91$ & $1.19$ & $0.2218$ & $0.0032$ & $6514.78$ & $9.6$ \\
$58727.86$ & $-2.92$ & $1.19$ & $0.2223$ & $0.0035$ & $6520.8$ & $10.4$ \\
$58728.71$ & $-4.13$ & $1.11$ & $0.2305$ & $0.0026$ & $6511.39$ & $7.3$ \\
$58728.76$ & $-3.05$ & $1.03$ & $0.2284$ & $0.0021$ & $6513.01$ & $5.9$ \\
$58729.76$ & $-5.6$ & $1.16$ & $0.2246$ & $0.0031$ & $6511.48$ & $9.0$ \\
$58729.86$ & $-5.54$ & $1.1$ & $0.226$ & $0.0029$ & $6511.24$ & $8.2$ \\
$58730.74$ & $-0.72$ & $1.49$ & $0.2145$ & $0.0051$ & $6510.23$ & $16.0$ \\
$58730.82$ & $1.18$ & $1.49$ & $0.2139$ & $0.0055$ & $6515.79$ & $17.1$ \\
$58744.77$ & $-1.12$ & $1.68$ & $0.2143$ & $0.0068$ & $6522.73$ & $21.2$ \\
$58745.74$ & $-16.3$ & $3.27$ & $0.2023$ & $0.0179$ & $6534.68$ & $61.2$ \\
$58746.69$ & $0.09$ & $1.54$ & $0.2148$ & $0.0055$ & $6517.99$ & $17.2$ \\
$58747.68$ & $6.78$ & $2.19$ & $0.2099$ & $0.01$ & $6505.19$ & $30.9$ \\
$58747.7$ & $4.5$ & $1.73$ & $0.2089$ & $0.0068$ & $6513.01$ & $22.0$ \\
$58748.75$ & $7.97$ & $1.56$ & $0.2267$ & $0.0056$ & $6507.91$ & $16.2$ \\
$58750.76$ & $-2.32$ & $1.4$ & $0.2289$ & $0.0046$ & $6522.1$ & $13.1$ \\
$58760.76$ & $-1.13$ & $1.79$ & $0.1976$ & $0.008$ & $6507.77$ & $28.3$ \\
$58760.85$ & $1.43$ & $1.34$ & $0.2064$ & $0.005$ & $6511.14$ & $16.5$ \\
$58761.72$ & $-1.0$ & $1.22$ & $0.2259$ & $0.0037$ & $6515.05$ & $10.7$ \\
$58762.71$ & $-2.72$ & $1.33$ & $0.2152$ & $0.0042$ & $6512.07$ & $13.1$ \\
$58763.78$ & $3.58$ & $1.13$ & $0.2196$ & $0.003$ & $6506.89$ & $9.1$ \\
$58764.67$ & $0.32$ & $1.21$ & $0.219$ & $0.0031$ & $6509.88$ & $9.5$ \\
$58765.71$ & $-43.31$ & $4.43$ & $0.23$ & $0.0271$ & $6569.43$ & $76.2$ \\
$58766.69$ & $-13.89$ & $1.11$ & $0.2193$ & $0.0026$ & $6502.88$ & $7.7$ \\
$58773.73$ & $-2.55$ & $1.66$ & $0.202$ & $0.0069$ & $6513.39$ & $23.6$ \\
$58774.69$ & $-10.79$ & $1.31$ & $0.2135$ & $0.0039$ & $6507.43$ & $12.3$ \\
$58774.8$ & $-13.66$ & $1.57$ & $0.1942$ & $0.0062$ & $6509.0$ & $22.5$ \\
$58775.67$ & $-10.02$ & $1.31$ & $0.2203$ & $0.0042$ & $6507.57$ & $12.5$ \\
$58775.81$ & $-11.1$ & $1.89$ & $0.2004$ & $0.0084$ & $6519.12$ & $29.1$ \\
$58776.74$ & $-0.04$ & $2.33$ & $0.2159$ & $0.0118$ & $6514.25$ & $36.6$ \\
$58776.8$ & $-8.42$ & $2.18$ & $0.2083$ & $0.011$ & $6509.26$ & $36.0$ \\
$58777.66$ & $1.5$ & $1.65$ & $0.2212$ & $0.0065$ & $6519.42$ & $19.3$ \\
$58777.78$ & $-4.05$ & $2.79$ & $0.2323$ & $0.015$ & $6512.9$ & $41.6$ \\
$58791.6$ & $-8.57$ & $1.61$ & $0.2249$ & $0.0061$ & $6511.53$ & $17.7$ \\
$58791.78$ & $-6.63$ & $1.85$ & $0.1908$ & $0.0084$ & $6506.83$ & $31.5$ \\
$58792.62$ & $-1.06$ & $1.31$ & $0.2249$ & $0.004$ & $6506.84$ & $11.6$ \\
$58792.76$ & $-2.15$ & $1.28$ & $0.2163$ & $0.0044$ & $6506.41$ & $13.5$ \\
$58793.62$ & $5.46$ & $1.26$ & $0.2296$ & $0.0033$ & $6508.11$ & $9.3$ \\
$58793.78$ & $6.2$ & $1.22$ & $0.2152$ & $0.004$ & $6510.24$ & $12.5$ \\
$58794.6$ & $2.03$ & $1.36$ & $0.2122$ & $0.0041$ & $6510.28$ & $13.1$ \\
$58794.71$ & $2.13$ & $1.3$ & $0.2181$ & $0.004$ & $6507.05$ & $12.2$ \\
$58795.59$ & $-2.05$ & $1.53$ & $0.2051$ & $0.0046$ & $6506.22$ & $15.3$ \\
$58795.72$ & $-1.34$ & $1.28$ & $0.213$ & $0.0037$ & $6506.98$ & $11.6$ \\
$58796.6$ & $-1.28$ & $1.17$ & $0.2217$ & $0.0029$ & $6505.65$ & $8.7$ \\
$58796.73$ & $-0.69$ & $1.2$ & $0.2181$ & $0.0034$ & $6502.47$ & $10.3$ \\
$58797.6$ & $-2.84$ & $1.31$ & $0.2107$ & $0.0036$ & $6503.46$ & $11.6$ \\
$58797.73$ & $-0.47$ & $1.41$ & $0.2095$ & $0.0046$ & $6507.76$ & $15.0$ \\
$58798.67$ & $-5.87$ & $1.21$ & $0.2189$ & $0.003$ & $6503.02$ & $9.2$ \\
$58799.75$ & $-16.63$ & $1.53$ & $0.203$ & $0.0058$ & $6509.75$ & $19.8$ \\
$58799.86$ & $-23.92$ & $1.56$ & $0.1928$ & $0.0064$ & $6525.7$ & $23.5$ \\

\hline
\hline
\end{tabular}
\end{table}

\begin{table}
\caption{HARPS spectroscopy obtained between 20th November 2019 and 17th January 2020.}
\label{tab:spec1}
\scriptsize
\begin{tabular}{lcccccc}
\hline
\hline
Time & RV & $\sigma_{\rm RV}$ & $S_{\rm MW}$ & $\sigma_{S}$ & FWHM & $\sigma_{\rm FWHM}$ \\
$[\rm{BJD}-2400000]$ & \multicolumn{2}{c}{[m/s]} & \multicolumn{2}{c}{--} & \multicolumn{2}{c}{[m/s]} \\
\hline
\hline

$58808.67$ & $-0.17$ & $1.29$ & $0.2389$ & $0.0041$ & $6538.41$ & $10.9$ \\
$58810.65$ & $4.55$ & $1.29$ & $0.2362$ & $0.0038$ & $6530.36$ & $10.4$ \\
$58811.71$ & $-0.94$ & $1.27$ & $0.228$ & $0.004$ & $6527.1$ & $11.3$ \\
$58816.73$ & $-7.0$ & $1.1$ & $0.2152$ & $0.003$ & $6504.0$ & $9.4$ \\
$58819.7$ & $-6.58$ & $1.4$ & $0.2183$ & $0.0044$ & $6501.38$ & $13.3$ \\
$58821.75$ & $-3.34$ & $1.41$ & $0.2111$ & $0.0047$ & $6496.4$ & $14.9$ \\
$58823.73$ & $-12.12$ & $1.5$ & $0.2006$ & $0.0053$ & $6497.09$ & $18.5$ \\
$58830.59$ & $-3.48$ & $0.98$ & $0.2136$ & $0.002$ & $6495.06$ & $6.2$ \\
$58830.69$ & $-4.1$ & $1.05$ & $0.2117$ & $0.0027$ & $6496.99$ & $8.7$ \\
$58831.54$ & $-7.06$ & $1.02$ & $0.2134$ & $0.0019$ & $6490.0$ & $6.1$ \\
$58831.67$ & $-9.5$ & $1.15$ & $0.2176$ & $0.0033$ & $6498.19$ & $9.9$ \\
$58832.55$ & $-10.45$ & $1.63$ & $0.2077$ & $0.0057$ & $6500.63$ & $18.7$ \\
$58832.66$ & $-9.7$ & $1.75$ & $0.2083$ & $0.0069$ & $6509.54$ & $20.8$ \\
$58833.61$ & $-1.92$ & $2.39$ & $0.1895$ & $0.0109$ & $6499.71$ & $41.1$ \\
$58834.58$ & $9.53$ & $1.31$ & $0.2155$ & $0.0039$ & $6500.35$ & $12.2$ \\
$58834.68$ & $9.52$ & $1.28$ & $0.2154$ & $0.0042$ & $6505.94$ & $12.9$ \\
$58835.58$ & $11.45$ & $1.06$ & $0.2275$ & $0.0022$ & $6514.18$ & $6.4$ \\
$58835.69$ & $8.87$ & $1.2$ & $0.2284$ & $0.0038$ & $6507.52$ & $10.7$ \\
$58836.55$ & $7.75$ & $1.05$ & $0.2285$ & $0.0022$ & $6518.42$ & $6.2$ \\
$58836.69$ & $7.74$ & $1.13$ & $0.2274$ & $0.0031$ & $6516.43$ & $8.8$ \\
$58837.57$ & $10.35$ & $1.23$ & $0.2329$ & $0.0034$ & $6519.22$ & $9.4$ \\
$58837.69$ & $8.19$ & $1.42$ & $0.2273$ & $0.0051$ & $6528.12$ & $14.6$ \\
$58838.66$ & $12.94$ & $1.33$ & $0.2383$ & $0.0044$ & $6522.37$ & $11.8$ \\
$58839.55$ & $9.99$ & $1.12$ & $0.2455$ & $0.0025$ & $6531.33$ & $6.4$ \\
$58839.61$ & $9.62$ & $1.14$ & $0.2524$ & $0.0029$ & $6533.14$ & $7.1$ \\
$58840.63$ & $4.6$ & $1.09$ & $0.2506$ & $0.003$ & $6536.72$ & $7.4$ \\
$58841.68$ & $9.26$ & $1.15$ & $0.243$ & $0.0035$ & $6536.59$ & $9.1$ \\
$58842.63$ & $14.97$ & $1.21$ & $0.2412$ & $0.0038$ & $6537.23$ & $9.9$ \\
$58843.56$ & $10.7$ & $1.03$ & $0.2515$ & $0.0021$ & $6535.97$ & $5.2$ \\
$58844.59$ & $0.13$ & $1.16$ & $0.2475$ & $0.0031$ & $6532.33$ & $7.8$ \\
$58845.67$ & $-4.45$ & $1.34$ & $0.2389$ & $0.0046$ & $6523.74$ & $12.3$ \\
$58847.65$ & $-8.6$ & $1.22$ & $0.2252$ & $0.0038$ & $6512.18$ & $11.1$ \\
$58848.6$ & $-12.47$ & $1.39$ & $0.2146$ & $0.0046$ & $6510.43$ & $14.4$ \\
$58848.61$ & $-13.08$ & $1.44$ & $0.2291$ & $0.0047$ & $6510.57$ & $13.4$ \\
$58849.64$ & $-9.39$ & $1.23$ & $0.219$ & $0.0038$ & $6507.37$ & $11.6$ \\
$58850.58$ & $0.44$ & $1.16$ & $0.2219$ & $0.0029$ & $6501.77$ & $8.4$ \\
$58852.57$ & $-3.63$ & $1.03$ & $0.22$ & $0.0024$ & $6503.71$ & $7.1$ \\
$58853.61$ & $-6.25$ & $1.18$ & $0.2178$ & $0.0034$ & $6498.99$ & $9.7$ \\
$58855.64$ & $2.87$ & $2.48$ & $0.1971$ & $0.0119$ & $6509.76$ & $42.5$ \\
$58865.55$ & $-5.45$ & $1.2$ & $0.2276$ & $0.0035$ & $6511.98$ & $9.9$ \\
\hline
\hline
\end{tabular}
\end{table}

\end{document}